\documentclass[
preprint,
showpacs,
amsmath,amssymb,
aps,
pre,
]{revtex4-1}

\usepackage{amsmath}
\usepackage{amsthm}

\usepackage{enumerate}
\usepackage{subfigure,wrapfig}

\usepackage{empheq}

\usepackage[dvipsnames]{xcolor}

\usepackage{graphicx}
\usepackage{dcolumn}
\usepackage{bm}
\usepackage{hyperref}
\usepackage[utf8]{inputenc}
\usepackage[english]{babel}
\usepackage[notref,notcite,final]{showkeys}

\RequirePackage[normalem]{ulem}
\RequirePackage{color}

\newcommand{\new}[1]{{\textcolor{black}{#1}}}

\begin{document}

\title{Bubble dynamics in a Hele-Shaw channel and velocity selection without surface tension}

			\author{Giovani L.~Vasconcelos}
		\email{giovani.vasconcelos@ufpr.br}%
		\affiliation{Departamento de F\'{\i}sica, Universidade Federal do Paran\'a, 81531-990 Curitiba, Paran\'a, Brazil}	
		\author{Luan P. Cordeiro}
		\affiliation{Departamento de F\'{\i}sica, Universidade Federal do Paran\'a, 81531-990 Curitiba, Paran\'a, Brazil}

		\author{Arthur A.~Brum}
		\affiliation{Departamento de F\'{\i}sica, Universidade Federal de Pernambuco, 50670-901, Recife, Brazil}
		\author{Mark Mineev-Weinstein}
		\email{mark\_mw@hotmail.com}%
		\affiliation{New Mexico Consortium, Los Alamos, NM, 87544, USA; \email{mark\_mw@hotmail.com}}

\begin{abstract}
\new{Inviscid bubble dynamics in a viscous fluid, moving with velocity $V$ far from the bubble, is considered. The Cauchy problem of recovering the bubble evolution from its initial shape is completely solved without surface tension. The well-posed (after a Tikhonov regularization) dynamical system provides an extensive list of {\it unsteady} closed form solutions due to integrability. A new (rational) class of solutions is obtained and added to the {\it logarithmic} class found earlier (Phys.~Rev.~E 89, 061003(R), 2014). The only attractor selects the observable bubble shape and velocity $U = 2V$  from the continuum of possible solutions. The attractor is asymptotically stable. Numerical results illustrate the most salient aspects of the bubble dynamics. }

\end{abstract}
	
	
\maketitle

\section{Introduction}

The motion of the interface between two fluids with a large viscosity contrast in a Hele-Shaw cell---a system where the fluids are confined between two closely spaced glass plates---has attracted considerable attention in both physics and mathematics. 
In physics, the Hele-Shaw model \cite{Lamb} is analogous to several other  systems, such as flows in porous media \cite{Darcy,Porous}, dendritic solidification \cite{Langer}, combustion fronts \cite{combustion}, electromigration of voids \cite{void}, streamer ionization fronts \cite{streamer}, and bacterial colony growth \cite{bacteria}, which are all governed (under certain approximations) by similar equations of motion \cite{Pelce}. In mathematics,   deep connections were revealed between Hele-Shaw flows and several apparently unrelated areas, such as 2D quantum gravity \cite{MWZ}, integrable systems \cite{KMWZ}, random matrices \cite{MPT}, quantum Hall effect \cite{ABWZ}, and Loewner evolution  \cite{vasiliev}. As a result, Hele-Shaw flows have become a fertile setting to study interface dynamics, pattern formation, and many other  problems in physics, mathematics, and applications.

One particular aspect of interface dynamics in a Hele-Shaw cell that has attracted a great deal of attention is the  {\it pattern selection problem}, posed in \cite{ST} in the context of viscous fingering. It concerns the question as to why the inviscid finger, moving inside a viscous fluid, reaches a relative width $\lambda=L_f/L_c=1/2$, where $L_f$ and $L_c$ are the finger and cell  widths, respectively. This implies that the finger velocity $U$ is twice the velocity $V$ of the background fluid, i.e., $U=2V$, despite the fact that the problem admits a continuum of stationary solutions with $0<\lambda<1$. In the mid-1980s, it was shown \cite{SurfaceTension} through a beyond-all-orders asymptotic analysis that the inclusion of surface tension leads to a discrete  (rather than continuous) spectrum of steady solutions, all of which converge to $U=2V$ 
in a zero surface tension limit. After that, numerical efforts have shown that only the lowest branch of the discrete spectrum is stable and therefore observable.
 Similar analysis was performed for the case of a single bubble in a Hele-Shaw channel  \cite{tanveer86,tanveer87}, where the velocity $U=2V$ was again selected.  

\new{Prior to addressing an alternative point of view to the selection problem, we should mention that the Hele-Shaw interface dynamics
without surface tension (in which case the problem is also known as {\it Laplacian growth}) is particularly fruitful and possesses remarkable properties untypical for most of nonlinear dynamical systems, because it is integrable.  
Steady exact solutions 
were found without surface tension:  for a finger \cite{ST}, for a single bubble \cite{TS} and for more complicated patterns --- the complete list of steady state solutions is in Ref.~\cite{GLV2015}. 
As for unsteady solutions, they are not possible to obtain unless the equations are integrable. Because of integrability, starting from \cite{Saffman1959}, several classes of unsteady solutions have been constructed   for the growth of fingers in a channel  \cite{Howison,Mineev94} and for an expanding bubble in an unbounded cell \cite{ShraimanBensimon, BensimonPelce,Howison, MarkSilvina}. (Refs.~\cite{Saffman1959, Howison,Mineev94, BensimonPelce, MarkSilvina}  present {\it regular} unsteady solutions which are valid for all times.) }


These unsteady exact solutions helped to discover in 1998 \cite{98PRL} that there is no need for surface tension for velocity  selection, as it turned out that the problem can be addressed  within the zero surface tension framework. Indeed, using a general class of exact time-dependent solutions that remain non-singular for all times, it was shown in Ref.~\cite{98PRL} that the steady finger with $U=2V$ is the only attractor of this dynamical system.
Subsequently, it was shown that the same selection mechanism (without surface tension) holds for a single bubble in a Hele-Shaw channel \cite{us2014}. Similar result was also obtained for a bubble in an unbounded Hele-Shaw cell  \cite{Robb2015}, where it was found that the circle (which has $U=2V$) is the only stable solution. The effect of surface tension on one and two bubbles in an unbounded cell \cite{Green2017} was  also studied and again the selected velocity $U=2V$ was obtained, in agreement with the predictions based on the zero-surface-tension dynamics. \new{The latest result on the subject \cite{PhysicaD2023} demonstrates that the law $U=2V$ also holds for an arbitrary number of inviscid bubbles in a Hele-Shaw channel.}

This accumulated body of work demonstrated that the special nature of the solutions with $U=2V$ is already “built-in” in the zero surface-tension dynamics, which is confirmed by the inclusion of regularization effects. This explains in part why the same selected velocity $U=2V$ is observed for other regularizing boundary conditions, such as kinetic undercooling \cite{mccue,king}, and even in non-fluid systems, such as finger-like ionization fronts in electric breakdown \cite{streamer}, where there is no analog of surface tension. In this context, constructing time-dependent solutions for Laplacian growth is important both mathematically, as they contribute to our understanding of nonlinear dynamics and integrable systems, and physically, since they help to describe and explain universal patterns observed in real systems.


\new{In this paper we present a detailed description of the unsteady motion of a single bubble in a Hele-Shaw channel in the absence of surface tension. 
Here are our main goals: (i) to present a new class of {\it  regular} solutions in terms of moving simple poles, which we refer to as rational solutions (in \cite{us2014} only logarithmic singularities were addressed); 
(ii) to blend both logarithmic and rational classes of unsteady solutions into a full single bubble dynamics; (iii) to address the velocity selection problem in this general context (rather than only for the logarithmic solutions as reported in \cite{us2014}); and (iv) to provide extensive numerical results to elucidate particularly salient features of the process.}


\new{Besides, the article contains the following results: i) we derive the symmetry properties that the problem possesses (whereas our exact solution reported in \cite{us2014} was constructed essentially by trial and error); and ii) on the basis of these symmetries we then show that in the long-time asymptotic regime any time-dependent solution  that exists for all time will (regardless of its specific form) reach a steady state  described by the Taylor-Saffman solution for a steadily moving bubble \cite{TS}. }


\new{It is well known  that zero-surface-tension exact solutions in Laplacian growth can either 
blow up in finite time (by cusp formation  or loss of univalence)  or stay regular \cite{Pelce, RMP1986,vasiliev}. 
Because of this feature, the zero surface tension problem is ill-posed in the {\it Hadamard sense} \cite{hadamard}. 
While many  problems in  dynamical systems  are found to be ill-posed in this sense, they still faithfully describe important physical processes \cite{lavrentev}.
However, as we indicated in \cite{PhysicaD2023}, this kind of ill-posed problems is usually dealt with by the Tikhonov's regularization  \cite{tikhanov}, whereby the initial  data  are restricted to those leading to regular solutions. Those initial data left after such elimination are called the {\it set of well-posedness} of the problem. This demonstrates that the mathematical results reported here are well-posed after Tikhonov's regularization and therefore correctly address  observable interface dynamics (in the limit of small surface tension).}

\new{Because the inviscid bubble is a `hole' surrounded by viscous fluid, 
the  flow domain is doubly connected.  So we introduce a conformal mapping from an annulus in an auxiliary complex $\zeta$-plane to the flow domain $D_t$ in the physical plane, $z=x+iy$; see below.} The corresponding mapping function $z(\zeta,t)$, where $t$ is time, is written explicitly in terms of the Jacobi Theta functions and possesses only logarithmic singularities or simple poles, whose locations vary in time.
The motion of these singularities in the $\zeta$-plane obeys a set of conserved quantities, which are {\it time-independent} singularities of the Schwarz function ${\cal S}(z,t)$ \cite{Davis} of the bubble interface; see below for a definition of ${\cal S}(z,t)$.
(Exact unsteady solutions for Hele-Shaw flows in other multiply connected geometries  have also been obtained both in an unbounded cell \cite{Richardson1994,Richardson1996a,Crowdy2002,pof2012,Jonathan2016c} and in the channel geometry \cite{Richardson1996b,Richardson2001,CrowdyTanveer,PhysicaD2023}, including the problem of flows past obstacles  \cite{Jonathan2016a, Jonathan2016b}.)

Existence of such conservation laws is due to the integrable nature of the Laplacian growth model \cite{MWZ}. The infinite-dimensional problem of interface motion in a Hele-Shaw cell is thus reduced to a finite-dimensional dynamical system, sometimes referred to as ``pole dynamics'' \cite{ShraimanBensimon, RMP1986},  which describes a motion of a finite number of singularities of the mapping $z(\zeta,t)$.
Exploring the properties of this dynamical system, we prove that under generic 
initial conditions \new{(after Tikhonov's regularization)} the bubble will approach a steady regime with velocity $U=2V$. 
More precisely, we show that if one starts with any initial shape whose Schwarz function ${\cal S}(z,0)$ is regular at infinity, then  $U=2V$ is always obtained; whereas initial shapes for which ${\cal S}(z,0)$ has a simple pole at infinity yield steady bubbles with $U\ne 2V$ (with $U$ being determined by the residue of the pole at infinity). However, any perturbation that removes the pole from infinity will  lead to $U=2V$, thus showing that this is the selected velocity in real systems (which are inevitably subjected to  perturbations, such as surface tension or other effects).
The selection scenario described above is akin to the {dynamical mechanism} for velocity selection \cite{Shraiman1984,Saarlos1988,Saarlos2003} observed in systems possessing traveling-wave solutions with different velocities, where most of the natural (generic) initial conditions lead to the same asymptotic velocity. Similarly, in our case, the behavior of the Schwarz function  at infinity determines the asymptotic velocity, with solutions with $U=2V$ being the only attractor of the dynamical system.


The present paper is organized as follows. In Sec.~\ref{sec:problem} we formulate the problem of the motion of a bubble in a Hele-Shaw channel. Secs.~III and IV contain all steady and unsteady single bubble solutions mentioned above. In Sec.~V we present several numerical examples of logarithmic solutions, both  regular and those losing univalence in finite time. In Sec.~VI we show  numerical examples for unsteady rational solutions.
In  Sec.~\ref{sec:selection} we perform a stability analysis that shows that the steady solutions with $U=2V$ present the only attractor of the dynamical system. Our results and conclusions are summarized in Sec.~\ref{sec:conclusions}.

\section{Formulation of the problem}  
 \label{sec:problem}
 
\begin{figure}[t]
\centering 
\subfigure[\label{fig:1a}]{\includegraphics[width=0.4\textwidth]{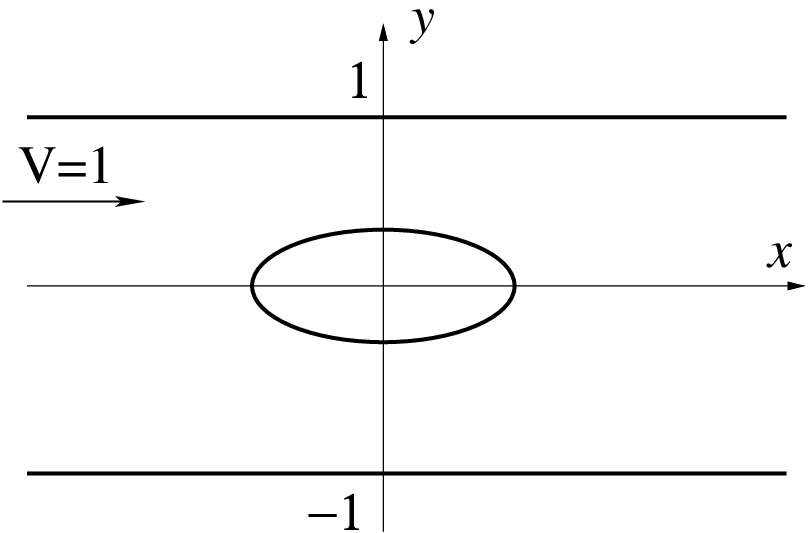}}  \qquad
\subfigure[\label{fig:1b}]{\includegraphics[width=0.4\textwidth]{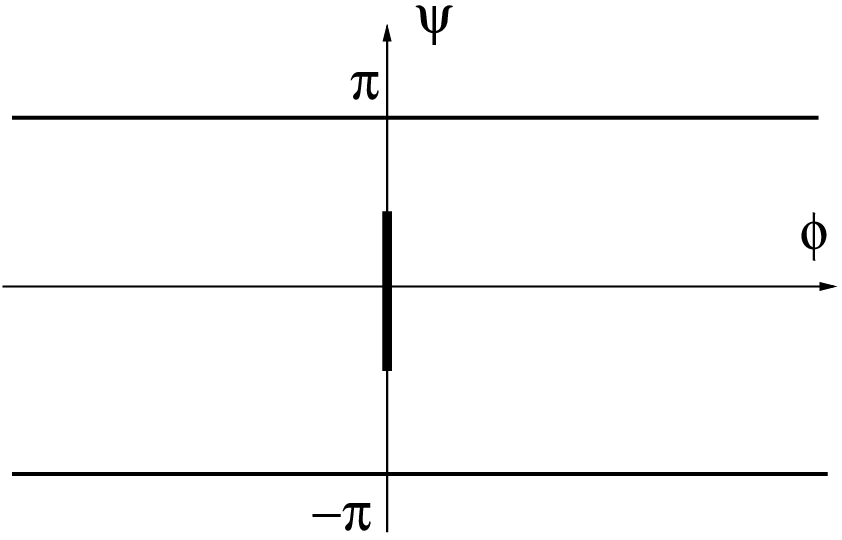}} \\
\subfigure[\label{fig:1c}]{\includegraphics[width=0.38\textwidth]{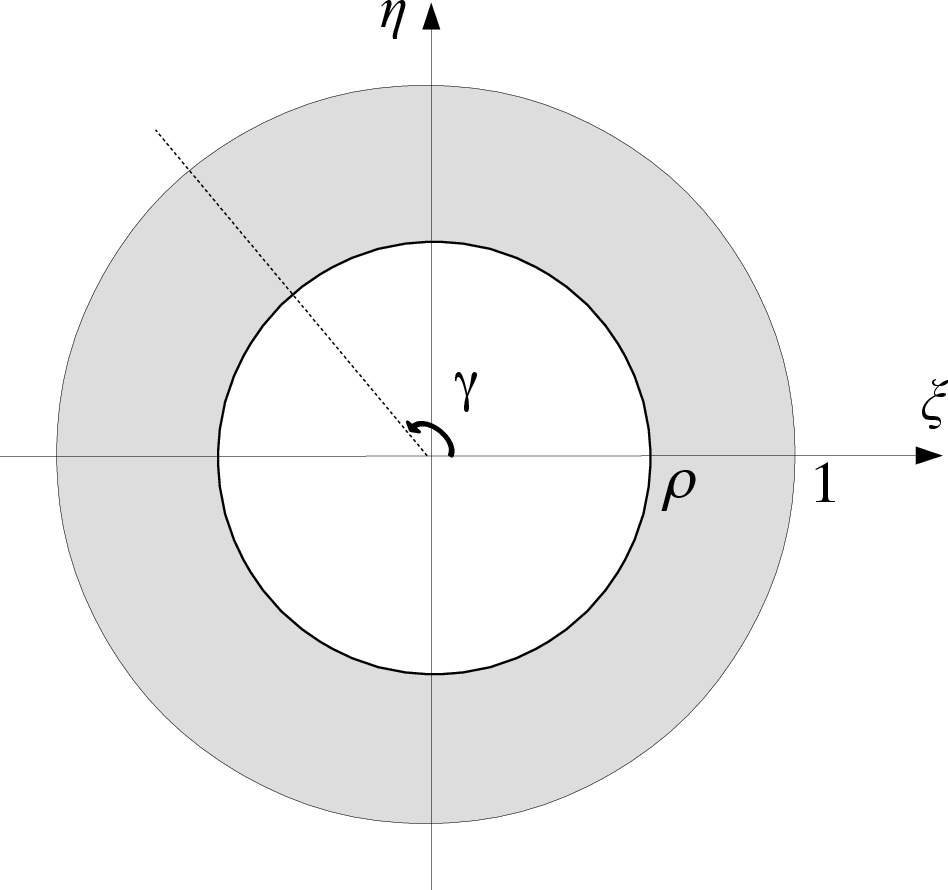}} \qquad
\subfigure[\label{fig:1d}]{\includegraphics[width=0.4\textwidth]{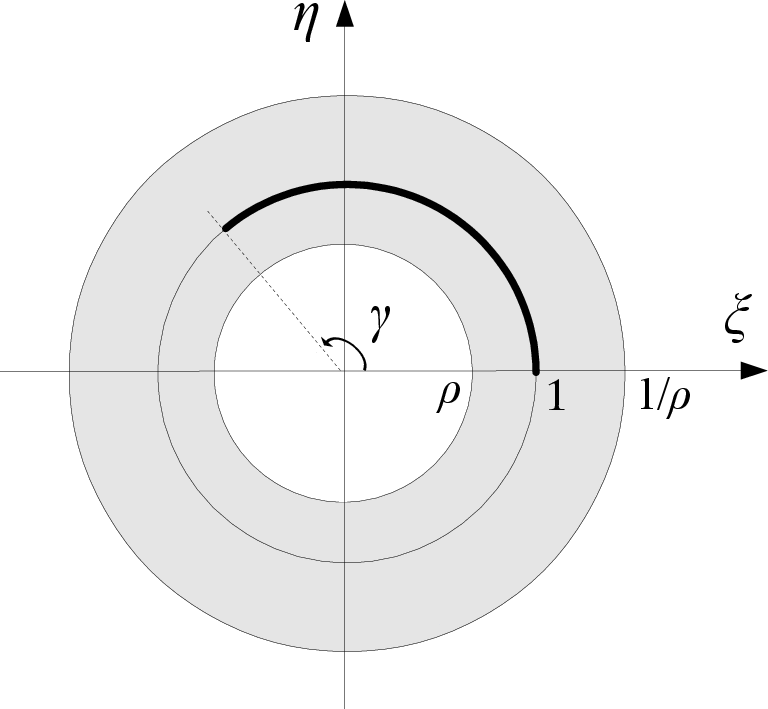}} 
\caption{The flow domains for a single bubble in  a Hele-Shaw channel: (a)  the physical $z$-plane; (b)  the plane of the complex potential;  (c)  the auxiliary complex $\zeta$-plane; and (d)  the extended domain in the  $\zeta$-plane.}
\label{fig:1}
\end{figure}

 We consider the problem of a single bubble of a less viscous fluid (say, air) moving in a more viscous fluid (say,  oil) inside a Hele-Shaw channel, which consists of two parallel rectangular glass plates separated by a small gap, $b$. The viscous fluid domain (outside the bubble) is denoted by $D(t)$, while the bubble boundary is denoted by $\partial D_t$.  The fluid inside the bubble is considered to be inviscid and is assumed to be at a constant pressure $p_0=0$. The viscous fluid is injected at a uniform rate by a source located at $x = -\infty$ and removed by a sink located at $x = +\infty$, thus generating a uniform flow with velocity $V$ in the far-field from the interface. Here we choose for convenience the channel width to be $\pi$ and set the far-field velocity to unity, i.e., $V=1$; see Fig.~\ref{fig:1a} for a schematics of the flow geometry.

Under the usual approximations \cite{Pelce}, the dynamics in a Hele-Shaw cell is governed by Darcy's law, ${\bf v} = - \frac{b^2}{12\mu} \nabla p$, where ${\bf v}(x,y,t)$ is the velocity vector field, $p(x,y,t)$ is the pressure field, $b$ is the cell gap, and $\mu$ is the fluid viscosity. Hele-Shaw flows are thus potential flows,  i.e., ${\bf v} =  \nabla \phi $, with  the velocity potential being  $\phi=- \frac{b^2}{12\mu}  p $. 
Assuming that the viscous fluid is incompressible, $\nabla \cdot {\bf v} = 0$, we find that  $\phi$ obeys the 2D Laplace equation, $\nabla^2 \phi = 0$.  Because of the uniform flow at infinity, we have $\phi = x$ for $x \to \pm \infty$. Also, $\partial_n \phi = 0$ (no normal flow) at both  channel walls, at $y = 0$ and $y=\pi$, respectively, where $\partial_n$ denotes the normal derivative. Furthermore, we shall neglect surface tension effects so that $p=0$ along the moving interface  $\partial D_t$, which implies that $\phi=0$ on $\partial D_t$.  Finally, the kinematic boundary condition requires that the normal velocities of the moving boundary, $V_n$, and of the viscous fluid, $v_n$, coincide, so $V_n = \partial_n \phi$ at the moving interface $\partial D_t$. The full mathematical formulation of the problem thus takes the following form:
\begin{subequations}
\label{eq:1}
\begin{align}
	& \nabla^2 \phi = 0 \qquad  \mbox{in}\qquad  D_t ,  \label{eq:1a}\\
	 &  \phi=0     \qquad \mbox{at}\qquad   \partial D_t ,  \label{eq:1b} \\
	 & \partial_n \phi = V_n  \quad \mbox{at}\qquad   \partial D_t, \label{eq:1c} \\
	 & \phi = x  \qquad \mbox{for}\qquad  x \to  \pm \infty,  \label{eq:1d}\\
	 & \partial_n \phi = 0  \qquad \mbox{at}\qquad  y = 0\ {\rm and}\  y=\pi.\label{eq:1e}
\end{align}
\end{subequations}

\new{In the model above we have neglected 
effects of the thin film coating left behind the moving inviscid pattern. There are regimes in viscous fingering where such effects  are relevant; see \cite{Tanveer1990} and references therein. But the  analysis presented here is valid for regimes where thin film effects are negligible. Most (if not all) works referenced above are in the same regime where one can safely neglect such effects.
}

In the next section we shall reformulate the problem in terms of a conformal mapping  from a circular domain in an auxiliary complex plane to the physical plane.

\subsection{The conformal mapping}

We  seek a solution to the problem in the form of a conformal map $z(\zeta,t)$ from a circular domain $D_\zeta$ in an auxiliary complex $\zeta$-plane to the fluid region $D_z$ in the  $z$-plane. The domain $D_\zeta$ consists of an annulus between the unit circle $C_0$ and an inner circle $C_1$ of radius $0<\rho<1$; see Fig.~\ref{fig:1c}.  
The inner circle $C_1$ is chosen to map to the interface (bubble boundary); while the unit circle $C_0$ maps to the channel walls, with the point $\zeta_+=1$ mapping to $x=\infty$ and the point $\zeta_-=e^{i\gamma}$, $0<\gamma<2\pi$, mapping to $x=-\infty$. With this choice, the upper and lower arcs of the unit circle, i.e., $C_0^ +=\{e^ {i\theta}, 0 < \theta < \gamma\}$ and $C_0^ -=\{e^ {i\theta}, \gamma < \theta < 2\pi\}$, are respectively mapped to the upper and lower channel walls; see Fig.~\ref{fig:1}. 

The mapping function $z(\zeta,t)$ must have logarithmic singularities at $\zeta=\zeta_\pm$:
\begin{align}
z(\zeta,t)&\approx \mp\ln (\zeta - \zeta_\pm), \quad \mbox{for}\quad \zeta \to\zeta_\pm.
\label{eq:logz}
\end{align}
But except for these singularities, $z(\zeta,t)$ must be analytic in the fluid domain $D_\zeta$ and satisfy the following boundary condition:
\begin{align}
{\rm Im}\,[z(\zeta,t)]= 0 \quad \mbox{for}\quad \zeta\in C_0^-,
\label{eq:Imz}
\end{align}
which implies, in view of the logarithmic singularities at $\zeta=\zeta_\pm$, see (\ref{eq:logz}), that the other required boundary condition, namely ${\rm Im}\,[z(\zeta,t)]=  \pi$ for $\zeta\in C_0^+$, is also satisfied. 
Using that $\overline \zeta=1/\zeta$ for $\zeta \in C_0$,  we can recast   (\ref{eq:Imz}) as
\begin{align}
\overline{z}(1/\zeta,t)= z(\zeta,t), 
\label{eq:barz}
\end{align}
where $\overline{f}(\zeta,t)$ denotes the Schwarz conjugate of $f(\zeta,t)$ defined by $\overline f(\zeta,t)=\overline{f\left(\overline \zeta,t\right)}$.
Relation (\ref{eq:barz}) is valid for $\zeta\in C_0^-$ and elsewhere by analytic continuation (assuming the same branch of the logarithmic function is chosen).

It is convenient to consider an augmented flow domain in the $z$-plane consisting of the original channel plus its reflection in the real axis, so that there are now two bubbles  within the flow domain, which are the mirror reflection of one another with respect to the centerline (real axis) of the extended channel. In such extended domain the solution becomes periodic in the $y$ direction with period $2\pi$.  The extended domain in the $\zeta$-plane, to be denoted by  $F_0$, is obtained by adding to $D_\zeta$ its reflection in the unit circle $C_0$, denoted by $\varphi_0(D_\zeta)$, thus generating an extended annulus:
\begin{equation}
F_0=D_\zeta\cup\varphi_0(D_\zeta)=\{\zeta~|~\rho<|\zeta|<1/\rho\},
\end{equation}
as shown in Fig.~\ref{fig:1d}. 
The function $z(\zeta,t)$ thus maps $F_0$ to the extended channel in the $z$-plane defined above. 
 
 \subsection{The complex potential}

 As the velocity potential $\phi(x,y,t)$ is a harmonic function, it is convenient to introduce the complex potential $w(z,t)=\phi(x,y,t)+i\psi(x,y,t)$, where $z=x+iy$ and $\psi(x,y,t)$ is  the stream function.  
It then follows from (\ref{eq:1}) that the complex potential $w(z,t)$ must be analytic in $D(t)$ and satisfy the following boundary conditions:
\begin{subequations}
\label{eq:WfBC}
\begin{align}
w(z)&\approx z \quad \mbox{for}\quad x \to\pm\infty,
\label{eq:wz}\\
{\rm Im}\,[w]&=0,\pi \quad \mbox{at}\quad y=0, \pi
\label{eq:Imw0}\\
{\rm Re}\,[w]&=0\quad\mbox{on}\quad  {\partial D_t}.
\label{eq:Rew}
\end{align}
\end{subequations}
From conditions (\ref{eq:Imw0})-(\ref{eq:Rew}), one sees that the flow domain in the $w$-plane corresponds to a horizontal strip, $0 < \psi < \pi$, $-\infty<\phi<\infty$, with a vertical slit inside it representing the bubble boundary, $\partial D(t)$, as shown in Fig.~\ref{fig:1b}.

Let us now introduce the function $W(\zeta,t)$ through the composition
\begin{align}
W(\zeta,t)\equiv w(z(\zeta,t)).
 \end{align}
The boundary conditions (\ref{eq:WfBC}) for the complex potential $w(z,t)$ can  now be recast in terms of the function $W(\zeta,t)$ as
\begin{subequations}
\label{eq:WfBCz}
\begin{align}
W(\zeta,t)&\approx \mp\ln (\zeta - \zeta_\pm) \quad \mbox{for}\quad \zeta \to\zeta_\pm,
\label{eq:wfz}\\
{\rm Im}\,[W]&= 0, \pi \quad \mbox{for}\quad \zeta\in C_0^\mp,
\label{eq:Imw0f}\\
{\rm Re}\,[W]&=0 \quad\mbox{for}\quad \zeta\in C_1.
\label{eq:Rewf}
\end{align}
\end{subequations}
Thus the mapping $w=W(\zeta,t)$ conformally takes the annulus $D_\zeta$ to the strip domain in the $w$-plane shown in Fig.~\ref{fig:1b}. 

In analogy to the argument leading from (\ref{eq:Imz}) to (\ref{eq:barz}), it follows that to fulfill (\ref{eq:Imw0f}) it suffices to require that
\begin{align}
\overline{W}(1/\zeta,t)&= W(\zeta,t).
\label{eq:barW1}
\end{align}
Similarly, using that $\overline \zeta=\rho^2/\zeta$ for $\zeta \in C_1$, condition (\ref{eq:Rewf})  can be recast as
\begin{align}
\overline{W}(\rho^2/\zeta,t)&= -W(\zeta,t),
\label{eq:barW2}
\end{align}
which is valid in $C_1$ and elsewhere by analytic continuation.
The symmetry relations (\ref{eq:barW1}) and (\ref{eq:barW2}) will be crucial to obtain the function $W(\zeta,t)$; see Sec.~\ref{sec:W}.

\subsection{The Schwarz function} 

The first three equations in (\ref{eq:1}) are known \cite{Howison92} to be equivalent to
\begin{equation}
{\cal S}_t = 2w_z,
\label{eq:St}
\end{equation}
where ${\cal S}(z,t)$ is the Schwarz function \cite{Davis} of the interface $\partial D_t$, defined as $\overline z = {\cal S}(z,t)$ for $z \in \partial D_t$. 
An important result follows from (\ref{eq:St}):  {\it all singularities of ${\cal S}(z,t)$ in $D(t)$  that are different from those of $w(z)$ must be constants of motion} \cite{Howison92}. As we will see below, this property is the key to the integrability of the Laplacian growth model.

Now, the Schwarz function of the interface $\partial D_t$  has the following useful representation in the $\zeta$-plane:
\begin{align}
	 g(\zeta,t)\equiv {\cal S}(z(\zeta,t),t) = \overline z(\rho^2/\zeta,t ).
		\label{eq:S}
\end{align}
where  we used that $\overline{\zeta}=\rho^2/\zeta$ for $\zeta\in C_1$. Notice that (\ref{eq:S}) is  valid for $\zeta \in C_1$ and elsewhere by analytic continuation. Furthermore, using (\ref{eq:barz}) in (\ref{eq:S}) gives
\begin{align}
	 g(\zeta,t) =  z(\zeta/\rho^2,t ).
		\label{eq:g2}
\end{align}
Next, using (\ref{eq:Imw0}) in (\ref{eq:St}) we conclude that 
${\rm Im}[{\cal S}(z)]=\mbox{constant}$ at  $y=0,\pi$, which 
 implies in turn that ${\rm Im}[g(\zeta,t)]=\mbox{constant}$ for $\zeta\in C_0$, or equivalently
\begin{align}
	 \overline {g}(1/\zeta,t) = g(\zeta,t )+2iC,
		\label{eq:barg}
\end{align}
for some real constant $C$.
Relation (\ref{eq:barg}) is valid on $C_0$ and elsewhere by analytic continuation.  Combining (\ref{eq:barg}) with (\ref{eq:S}) yields the following additional symmetry for the map $z(\zeta,t)$:
\begin{align}
	 z(\rho^2 \zeta,t) = z(\zeta/\rho^2,t ) +i2C.
	 \label{eq:z4}
\end{align}
or equivalently
\begin{align}
	 z(\rho^{\pm4} \zeta,t) = z(\zeta,t ) \pm i2C \ .
		\label{eq:zrho}
\end{align}
This relation is crucial in constructing exact solutions for $z(\zeta,t)$, as we will see later.

Before we discuss specific solutions, let us make a small technical digression. Let us introduce the map $\psi_1(\zeta)=\rho^4\zeta$ and its inverse $\psi_{-1}(\zeta)=\rho^{-4}\zeta$, and define the set $\Theta_1=\{\psi_n(\zeta)=\rho^{4n}\zeta|n\in\mathbb{Z}\}$  generated by iterations of the maps $\psi_{\pm 1}(\zeta)$.
One can check that the set $\Theta_1$ thus defined is an example of what is known as a classical {\it Schottky group}. 
More precisely, $\Theta_1$ is a subgroup of the `primary' Schottky group $\Theta=\{ \theta_{n}(\zeta)=\rho^{2n}\zeta |n\in\mathbb{Z}\}$ generated by the map $\theta_1(\zeta)=\rho^2\zeta$ and its inverse $\theta_{-1}(\zeta)=\rho^{-2}\zeta$, such that  $\Theta_1$  contains only an even number of compositions of the  maps $\theta_{\pm1}$.
Equation (\ref{eq:zrho}) then says that the mapping function $z(\zeta,t)$ is {\it automorphic} up to an additive factor with respect to the group $\Theta_1$, meaning that $z(\zeta,t)$ is invariant (up to an additive constant) under each element of $\Theta_1$:
\begin{align}
	 z(\psi_{n} (\zeta),t)= z(\zeta,t ) +\mbox{constant}.
  \label{eq:zsymm}
\end{align}


\section{Steady solutions}

\label{sec:steady}

Before proceeding to discuss time-dependent solutions, we shall first review  (within our notation) the known solutions \cite{TS,tanveer87,GLV2001} for a bubble {\it steadily} moving with speed $U$ in a Hele-Shaw channel.  Knowing the explicit form of these steady solutions is a crucial step in constructing time dependent solutions, as we will see in Sec.~\ref{sec:exact}.

\subsection{The complex potential}
\label{sec:W}

The complex potential satisfying the boundary conditions (\ref{eq:WfBCz}) is given by 
\begin{equation}
W(\zeta,t) = i\frac{\gamma}{2} +
\log\,\frac{P(e^{-i\gamma}\zeta;\rho^2) P(\rho^2\zeta; \rho^2)}{P(\zeta;\rho^2) P(\rho^2
e^{-i\gamma}\zeta;\rho^2)},
\label{eq:Wb}
\end{equation}
where the function $P(\zeta; \rho)$ is defined by
\begin{equation} P(\zeta; \rho) =
(1-\zeta) \prod_{m=1}^{+\infty}(1-\rho^{2m}\zeta)(1-\rho^{2m}/\zeta).
\label{eq:P}
\end{equation}
For later use we note that  $P(\zeta;\rho)$ satisfies the following symmetry relations:
\begin{gather}
P(\rho^2\zeta;\rho)=P(1/\zeta;\rho)=-\frac{1}{\zeta}P(\zeta;\rho)
\label{eq:P1}
\end{gather}
from which we can derive other useful relations:
\begin{gather}
 P(\rho/\zeta;\rho)=-\frac{\rho}{\zeta}P(\zeta/\rho;\rho)=P(\rho\zeta;\rho),
 \label{eq:P2}
\\
 P(\rho^2/\zeta;\rho)=  -\frac{\rho^2}{\zeta} P(\zeta/\rho^2;\rho)=  P(\zeta;\rho).  \label{eq:P4}
\end{gather}
From the definition of $P(\zeta;\rho)$ it is clear that $W(\zeta,t)$ defined in (\ref{eq:Wb}) has the desired logarithmic singularities indicated in (\ref{eq:wfz}). Furthermore, using the above symmetry relations for $P(\zeta;\rho)$,  it is easy to verify  that $W(\zeta,t)$  satisfies the  symmetry relations (\ref{eq:barW1}) and (\ref{eq:barW2}); details are left for the reader. We remark that the $P$-function defined above is related to the first Jacobi theta function, $\vartheta_1(z;\rho)$ \cite{Gradshtein}, through
\begin{equation} P(\zeta;q) =
\frac{\sqrt{\zeta} \, \vartheta_1(-i \ln \sqrt{\zeta};q)}{i q^{1/4}\prod_{m=1}^{\infty}(1-q^{2m})},
\end{equation}
where
\begin{equation*}
\vartheta_1(z;q) = 2q^{1/4}\sin z\prod_{m=1}^{\infty} \left(1-q^{2m}\right)\left(1-2q^{2m}\cos 2z+q^{4m}\right). 
\end{equation*}

\subsection{The conformal  mapping}

As shown in Ref.~\cite{us2014}, 
the conformal map $z_0(\zeta,t)$ that describes a steady bubble is given by 
\begin{align}
z_0(\zeta, t) = d(t) + i\frac{\gamma}{2}+ \log \frac{P(e^{-i\gamma}\zeta;\rho^2)}{P(\zeta;\rho^2)} 
 +   \alpha_0 \log\frac{P(\rho^2 e^{-i\gamma}\zeta; \rho^2)}{P(\rho^2\zeta;\rho^2)},
\label{eq:z0}
\end{align}
where $d(t)=Ut$, $\alpha_0$ is a real constant with $|\alpha_0| < 1$, and the bubble speed $U$ is given in terms of the parameter $\alpha_0$ by
\begin{align}
U=\frac{2}{1-\alpha_0}.
\label{eq:U}
\end{align}
In the steady solutions above the parameters $\gamma$ and $\rho$ are constant, whereas for unsteady solutions they become time dependent; see Sec.~\ref{sec:exact}.

Using (\ref{eq:g2}), (\ref{eq:P2}), and (\ref{eq:z0}), we obtain that the Schwarz function in the $\zeta$ plane associated with the map (\ref{eq:z0}) is
\begin{align}
g_0(\zeta, t) = d(t)- i \frac{\gamma}{2}+ \log\,\frac{P(\rho^2 e^{-i\gamma}\zeta;\rho^2) }{P(\rho^2\zeta ;\rho^2)}
+ \alpha_0 \log\frac{P(e^{-i\gamma}\zeta; \rho^2)}{P(\zeta;\rho^2)} .
\label{eq:g0}
\end{align}
Note that the only singularities of $g_0(\zeta,t)$ in the fluid domain $D_\zeta$ are at the points $\zeta=\zeta_\pm$ (recall that $\zeta_+=1$ and $\zeta_-=e^{i\gamma}$), which are the pre-images of $x=\pm\infty$. 

Now, comparing (\ref{eq:g0}) with (\ref{eq:z0}), one can show that the Schwarz function ${\cal S}_0(z,t)$ has the following asymptotic behaviour at infinity:
\begin{align}
{\cal S}_0(z,t) = \alpha_0 z + B_\pm^0 + O\left(\frac{1}{z}\right), \qquad \mbox{for}  \quad x\to \pm \infty,
\label{eq:Sz}
\end{align}
 where
\begin{align}
B_\pm^0=&  (1-\alpha_0) d(t) \pm
(1 -\alpha_0^2) \log \frac{P(\rho^2 e^{i\gamma};\rho^2)}{P(\rho^2;\rho^2)} - i\frac{\gamma}{2}\left(1+\alpha_0\right) .
\label{eq:C0pm}
\end{align}
If $\alpha_0\ne 0$ (i.e., $U\ne 2$)  then the Schwarz function ${\cal S}_0(z,t)$ has a simple pole at $|x|=\infty$, whereas for $\alpha_0=0$ (i.e., $U=2$)  ${\cal S}_0(z,t)$ is regular in the entire fluid domain. This property will be very important for the problem of velocity selection to be discussed later.

\section{Unsteady solutions: general properties}

\label{sec:exact}

\subsection{Conformal mapping} 

We shall consider the following general solution for the conformal mapping $z(\zeta, t)$ describing the  bubble dynamics:
 \begin{align}
z(\zeta, t) =~& z_0(\zeta,t)
 + h(\zeta,t),
\label{eq:zg}
\end{align}
where $z_0(\zeta,t)$ is as given in (\ref{eq:z0}) but where now $d(t)$ is a generic time-varying, real-valued parameter to be determined later. Also, the  parameters $\gamma$ and $\rho$ become time dependent, so that their time evolution needs to be determined as part of the solution. We note, however, that $\alpha_0$ remains constant. The general solution given in (\ref{eq:zg}) can  be seen as a generic perturbation (not necessarily small) of the steady solutions, where the initial bubble shape differs from the steady shape by the addition of the term $h(\zeta,0)$. For subsequent times the interface will evolve according to the mapping (\ref{eq:zg}).

Before we consider specific forms for the function  $h(\zeta,t)$, let us discuss some general properties that it must satisfy. First,  it is clear that $h(\zeta,t)$ must be analytic  in $D_\zeta$. Second, in view of the symmetry relations (\ref{eq:barz}) and (\ref{eq:zsymm}) it follows that
\begin{align}
\overline{h}(1/\zeta,t) = h(\zeta,t), 
\label{eq:barh}
\end{align}
and
\begin{align}
	 h(\rho^{4n} \zeta,t) = h(\zeta,t ) +\mbox{constant} 
  , \qquad n\in \mathbb{Z}.
		\label{eq:hrho}
\end{align}

Now suppose that $h(\zeta,t)$ has singularities, say, either logarithmic branch points or simple poles, at some pre-specified points $\{a_k|k=1,...,N\}$, where $a_k\notin F_0$. Then from (\ref{eq:barh})  it follows that the points $1/\overline{a_k}$ are also singular points (of the same nature) of $h(\zeta,t)$. This implies  in view of (\ref{eq:hrho}) that the full set of singularities of $h(\zeta,t)$ is $\{\rho^{4n}a_k, \rho^{4n}/\overline a_k\}$, for $n\in\mathbb{Z}$.
Furthermore, in view of (\ref{eq:g2}) we see that $g(\zeta,t)$ has singularities at the points $\{\rho^{2n}a_k, \rho^{2n}/\overline a_k\}$. Without loss of generality, let us assume 
that $\rho^2<|a_k(0)|<\rho$. From the preceding discussion, one can check that the only singularities of  $g(\zeta,t)$ in the fluid region $F_0$, besides those at $\zeta_\pm$,  are the points $\zeta_k=\rho^2/\overline{a}_k$ and $\varphi_0(\zeta_k)= a_k/\rho^2$, for $k=1,...,N$. This implies in turn that  ${\cal S}(z,t)$ has singularities inside the fluid domain $F_0$ at the points $\beta_k$ and $\overline{\beta}_k$, where
\begin{align}
\beta_k \equiv z(\rho^2/\overline{a}_k,t)=z_0(\rho^2/\overline{a}_k,t)+ h(\rho^2/\overline{a}_k,t),
\end{align}
or more explicitly
\begin{align}
 \beta_k = d(t) + i\frac{\gamma}{2}+ \log \frac{P(\rho^2e^{-i\gamma}/\overline{a}_k;\rho^2)}{P(\rho^2/\overline{a}_k;\rho^2)} 
 +   \alpha_0 \log\frac{P(e^{i\gamma}\overline a_k; \rho^2)}{P(\overline a_k;\rho^2)}+ h(\rho^2/\overline{a}_k,t).
\label{eq:betakg}
\end{align}
Since  the complex potential $w(z,t)$ is regular at these singular points  of ${\cal S}(z,t)$, it follows from the property outlined after (\ref{eq:St}) that the quantities $\beta_k$ must remain fixed in time:
\begin{align}
\dot \beta_k=0.
\label{eq:dotbetak}
\end{align}

Furthermore, not only the location $\beta_k$ of the singularities of ${\cal S}(z,t)$ must remain fixed, but so do their strength. Here there are two cases to consider. First, suppose that $h(\zeta,t)$ has a logarithmic singularity at $a_k$, that is,
\[ h(\zeta,t) \approx \alpha_k \log (\zeta-a_k), \qquad \mbox{as}\qquad \zeta\to a_k,\]
for some complex parameter $\alpha_k$. This implies turn  that ${\cal S}(z,t)\approx \alpha_k \log (z-\beta_k)$, for $z\to\beta_k$. Hence, $\alpha_k$ must be constant. In addition, it is required that 
\begin{align}
\sum_{k=1}^N \alpha_k = 0,
\label{eq:a0}
\end{align} 
to ensure univalence of $h(\zeta,t)$.

Similarly, if $a_k$ is a pole of $h(\zeta,t)$, that is,
\[ h(\zeta,t) \approx \frac{r_k}{\zeta-a_k} \qquad \mbox{as} \qquad \zeta\to a_k,\]
then 
the residue $R_k$ of ${\cal S}(z,t)$ at the corresponding pole $\beta_k$ must remain constant: 
 \begin{align}
\dot R_k=0,
\label{eq:dotRk}
\end{align}
where
\begin{align}
 R_k = -\frac{\rho^2\overline r_k}{\overline a_k^2}z_\zeta(\rho^2/\overline a_k).
 \label{eq:Rk}
 \end{align}
(Poles of higher order could in principle be also considered but for simplicity we shall not discuss this detail here.)

Three additional real conserved quantities can be found by analyzing the behavior of ${\cal S}(z,t)$ at infinity.
Proceeding as  in Sec.~\ref{sec:steady}, one can show that  ${\cal S}(z,t)$ has the following asymptotic behavior at infinity:
\begin{align}
{\cal S}(z,t) = \alpha_0 z + B_\pm + O\left(\frac{1}{z}\right), \qquad \mbox{for}  \quad x\to \pm \infty,
\label{eq:Szg}
\end{align}
 where
\begin{align}
B_\pm = B_\pm^0 +h(\zeta_\pm/\rho^2,t) -\alpha_0 h(\zeta_\pm,t) .
\end{align}
with $B_\pm^0$ as given in (\ref{eq:C0pm}).  Now inserting (\ref{eq:Szg}) into (\ref{eq:St}) and noting that $w_z\approx 1$ for $|z|\to \infty$, we then conclude that 
\[ \dot B_\pm = 2,\]
thus implying that the quantities $\beta_\pm$ defined by
\begin{align}
\beta_\pm =   B_\pm - 2t
\label{eq:betapm}
\end{align}
are constants of motion:
\begin{align}
\dot{\beta}_\pm=0.
\label{eq:dotbetapm}
\end{align}

More explicitly the  conserved quantities $\beta_\pm$ read
\begin{align}
\beta_\pm =    (1-\alpha_0) d(t) -2t \pm
(1 -\alpha_0^2) \log \frac{P(\rho^2 e^{i\gamma};\rho^2)}{P(\rho^2;\rho^2)} - i\frac{\gamma}{2}\left(1+\alpha_0\right) + h(\zeta_\pm/\rho^2,t) -\alpha_0 h(\zeta_\pm,t)
 \label{eq:betapmg}
\end{align}
Using the symmetry relations for $P(\zeta;\rho)$ and $h(\zeta,t)$, one can verify  that $\beta_\pm$ have the same imaginary part: ${\rm Im}[\beta_+]={\rm Im}[\beta_-]$.
This shows that there are only three real conserved quantities between the two complex constants $\beta_\pm$.

Summarizing our discussion thus far, we have seen that once the form of the function $h(\zeta,0)$ is given---restricted here to the class of functions possessing either logarithmic branch points or poles---, the time evolution of its singularities $a_k$ as well as of the parameters $d(t)$, $\gamma(t)$, and $\rho(t)$ are determined by the conserved quantities $\beta_k$ and $\beta\pm$,  whose fixed values are set by the initial conditions. 
Furthermore, if said singularities are logarithmic branch points, then their strength $\alpha_k$ are constant; whereas if they are poles, their residues $r_k$ evolve in time in such a way so as to keep the corresponding residue $R_k$ of the Schwarz function constant. As we will see next, the existence of this set of conserved quantities impose severe restriction of the possible asymptotic evolution of the singularities $a_k$ for $t\to\infty$.

\subsection{Long-time asymptotic regime.} 

\label{sec:asymptotic}

It follows from (\ref{eq:betapmg}) that in order to  keep $\beta_\pm$ constant for all times, the parameter $d(t)$ must behave as
\begin{align}
d(t)=Ut\qquad \mbox{for}\qquad t\to \infty,
\label{eq:d}
\end{align}
with $U$ as defined in (\ref{eq:U}). This follows from the fact that all other terms in the right-hand side of (\ref{eq:betapmg}) stay finite for all times.

Similarly, using  (\ref{eq:d}) in (\ref{eq:betakg}), one sees that a term in the right-hand side of (\ref{eq:betakg}) must cancel the divergence of $d(t)$ for $t\to\infty$, so as to ensure that $\beta_k$ remains constant for all times. Inspection of (\ref{eq:betakg}) reveals that the only such possibility is to have
\begin{align}
a_k\to\rho^2 e^{i\gamma}\qquad\mbox{for}\qquad t\to \infty.
 \label{eq:a_kinf}
 \end{align}
We thus conclude that all singularities $a_k$, regardless of whether they are log-branch points or poles,  must asymptotically converge to the same point $a_0^-=\rho^2 e^{i\gamma}$. Furthermore, one can show that in either case $h(\zeta,t)\to 0$ for $t\to\infty$. For instance, if  $h(\zeta,t)$ has only logarithmic singularities, they will all approach the same point and cancel off because of condition (\ref{eq:a0}), thus resulting in $h(\zeta,t)\to 0$ for $t\to\infty$.
Similarly, if  $h(\zeta,t)$ is a rational function with simple poles at  points $a_k$ with residue $r_k$, and $a_k\to\rho^2 e^{i\gamma}$, then condition (\ref{eq:Rk}) implies that $r_k\to0$ for $t\to\infty$, so as to cancel the divergence of the term $z_\zeta(e^{i\gamma})$ and  keep $R_k$ constant, which then implies that $h(\zeta,t)\to 0$ for $t\to\infty$.
We thus conclude that any time dependent solution that exists for all times 
will necessarily reach a steady regime described by the solution (\ref{eq:z0}), where the bubble propagates with a constant speed $U$ dictated by the parameter $\alpha_0$, as indicated in (\ref{eq:U}).

 
\section{Logarithmic Solutions}
\label{sec:log}


Here we shall consider the  following solution for the function $h(\zeta,t)$ \cite{us2014}:
\begin{align}
h(\zeta, t) = \sum_{k=1}^N \left\{ \alpha_k \log P(a_k/\zeta;\rho^2) + \overline \alpha_k \log P(\overline a_k \zeta;\rho^2)\right\},
\label{eq:hl}
\end{align}
where  $a_k(t) \notin D_\zeta$, $k=1,..,N$,  are complex time-dependent parameters and $\alpha_k$ are complex constants satisfying the condition (\ref{eq:a0}). As already mentioned, the initial values of $a_k(0)$ are chosen such that $\rho^2<|a_k(0)|<\rho$. It is clear  that by construction the function defined in (\ref{eq:hl})   satisfies (\ref{eq:barh}). Also, using relations (\ref{eq:P1}) and (\ref{eq:P2}), it is not difficult to check it also  has the additional required symmetry  indicated in (\ref{eq:hrho}).

The conserved quantities $\beta_k$ and $\beta_\pm$ defined respectively in (\ref{eq:betakg}) and (\ref{eq:betapmg}) become
 \begin{align}
\beta_k = & d(t) +i\frac{\gamma}{2}+ \log \frac{P(e^{-i\gamma}\rho^2/\overline a_k;\rho^2)}{P(\rho^2/\overline a_k;\rho^2)}  +   \alpha_0 \log\frac{P(e^{i\gamma}\overline a_k; \rho^2)}{P(\overline a_k;\rho^2)}
 + \cr
 +& \sum_{m=1}^N \left\{ \alpha_m \log P(a_m \overline a_k /\rho^2;\rho^2) + \overline \alpha_m \log P(\rho^2 \overline a_m /\overline a_k;\rho^2)\right\},
\label{eq:betakl}
\end{align}
and 
\begin{align}
\beta_\pm = &   (1-\alpha_0) d(t) -2t \pm
(1 -\alpha_0^2) \log \frac{P(\rho^2 e^{i\gamma};\rho^2)}{P(\rho^2;\rho^2)} - i\frac{\gamma}{2}\left(1+\alpha_0\right) +\cr 
&+ \sum_{k=1}^N \left\{  \alpha_k \log P(\rho^2a_k/\zeta_\pm;\rho^2) +  \overline \alpha_k \log P(\overline a_k \zeta_\pm/\rho^2 ;\rho^2) \right\}\cr
&- \alpha_0 \sum_{k=1}^N \left\{ \alpha_k \log P(a_k/\zeta_\pm;\rho^2) + \overline \alpha_k \log P(\overline a_k \zeta_\pm;\rho^2)\right\}.
 \label{eq:betapml}
\end{align}

As already discussed, equations (\ref{eq:betakl}) and (\ref{eq:betapml}) provide the complete time-dependence of the parameters $\gamma(t)$,  $\rho(t)$,  $d(t)$, and $a_k(t)$,  once their  initial values are given. More explicitly, if we write $a_k(t)=\xi_k(t)+ i\eta _k(t)$, we can  use the fact that $\dot\beta_k=0$ and $\dot \beta_\pm=0$ to obtain a set of $2N+3$ ordinary differential equations for equal number of real-valued parameters, namely: $d$, $\gamma$,  $\rho$, and $\{\xi_k, \eta _k|k=1,...N; \}$. By  integrating  these equations numerically, it is then possible to compute the time evolution of the interface.  Initial conditions should however be chosen carefully, since some of them may lead to a break-down of the solutions in finite time because of loss of univalence of the mapping $z(\zeta,t)$.
\new{(But these initial data are eliminated by Tikhonov's regularization mentioned above.)}
Below we show several examples of bubble evolution described by our solutions. Most of the initial data in these examples belong to the set of well-posedness (see above), but a couple of them outside this set are  included for illustrative purposes.

\begin{figure}[t!]
	\center \subfigure[\label{fig:b1a}]{\includegraphics[width=0.5\textwidth]{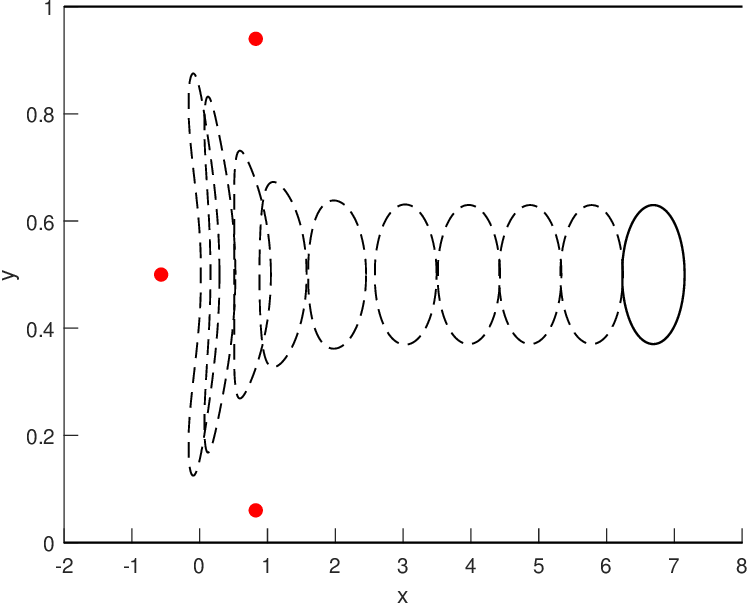}}\\ 
	 \subfigure[\label{fig:b1b}]{\includegraphics[width=0.35\textwidth]{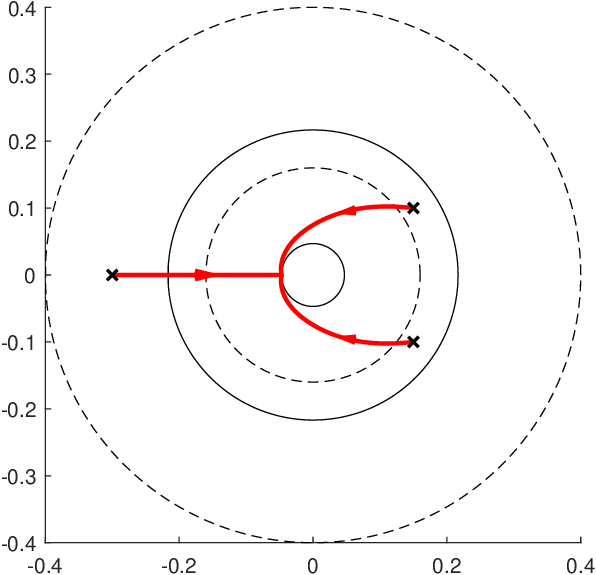}}\qquad\quad
	 \subfigure[\label{fig:b1c}]{\includegraphics[width=0.4\textwidth]{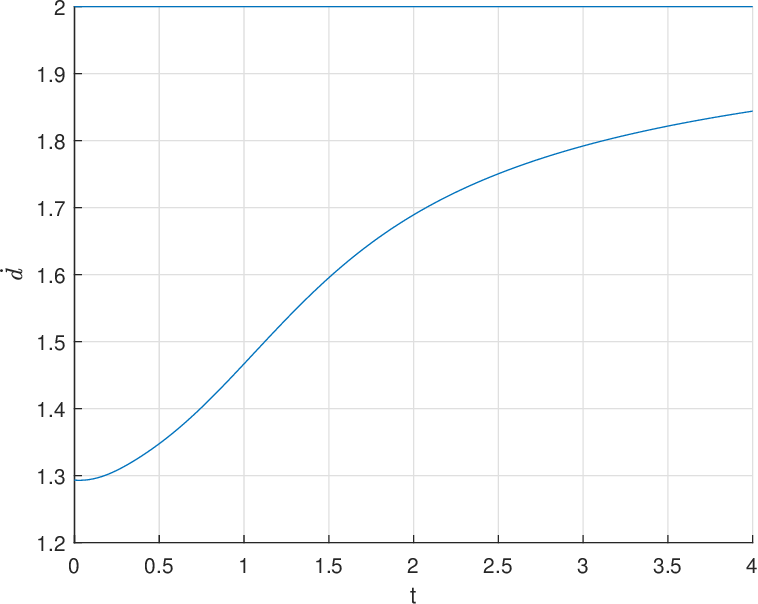}}
		\caption{(a) Bubble evolution leading to a symmetric steady shape with $U=2$. The initial shape corresponds to Eqs.~(\ref{eq:zg}) and (\ref{eq:hl}) with the parameters $\gamma=\pi$, $\alpha_1=-0.4$, $\alpha_2=\alpha_3=0.2$, $a_1(0)=-0.3$, $a_2(0)=\overline{a_3(0)}=0.15+0.1i$, and $\rho(0)=0.4$. Successive shapes are shown at the time instants $t=0, 0.1724, 0.5228, 0.8340, 1.3011, 1.8275, 2.2960, 2.7487, 3.2021, 3.6570$. The red dots indicate the locations of the singularities $\beta_k$, $k=1,2,3$, of the Schwarz function. (b) The trajectories (red curves) of the singularities   $a_k$ in the $\zeta$ plane, where the black crosses indicate their initial locations. Here the dashed (solid) circles correspond to the initial (final) circles of radii $\rho$ and $\rho^2$.   (c) The time derivative of the parameter ${d}(t)$ plotted as a function of time.}
	\label{fig:b1}
\end{figure}

As shown in Sec.~\ref{sec:steady}, if a solution exists for all times then the bubble will necessarily reach a steady regime where it moves with a constant speed $U$, whose value is determined by the parameter $\alpha_0$ via (\ref{eq:U}). Let us first consider solutions where the bubble attains the asymptotic velocity $U=2$, which occurs whenever  $\alpha_0=0$. 

In Fig.~\ref{fig:b1} we show an example of this case for a symmetric initial condition corresponding to the following parameters: $\gamma=\pi$, $\alpha_1=0.2$, $\alpha_2=-0.4$, $\alpha_3=0.2$, $a_1(0)=-0.3$, $a_2(0)=\overline{a_3(0)}=0.15+0.1i$. In this and all subsequent cases we take $\rho(0)=0.4$.  In Fig.~\ref{fig:b1a} we display  plots of the bubble interface at successive times. The red dots in this figure indicate the location of the fixed singularities $\beta_k$, $k=1,2,3$, of the Schwarz function ${\cal S}(z,t)$.  In Fig.~\ref{fig:b1b} we show the trajectories (red curves) of the three singularities $a_k$ in the $\zeta$ plane, where their initial locations are denoted by black crosses. (The conventions used in this figure will be adopted for the remaining figures.) The convergence of all singularities to the special point $a_0^-=-\rho^ 2$, for $t\to\infty$, is clearly visible in this figure. In Fig.~\ref{fig:b1c} we plot the quantity $\dot{d}$, which yields an estimate of the bubble velocity, as a function of time. Comparing Figs.~\ref{fig:b1a}  and \ref{fig:b1c}, one sees that the interface moves slowly at first, i.e., $\dot d (0)\approx 1.3<U=2$,  but once it gets past the two singularities of ${\cal S}(z,t)$ that are initially upstream of the bubble,  it  quickly speeds up toward the asymptotic velocity $U=2$.

\begin{figure}[t]
	\center \subfigure[\label{fig:2a}]{\includegraphics[width=0.4\textwidth]{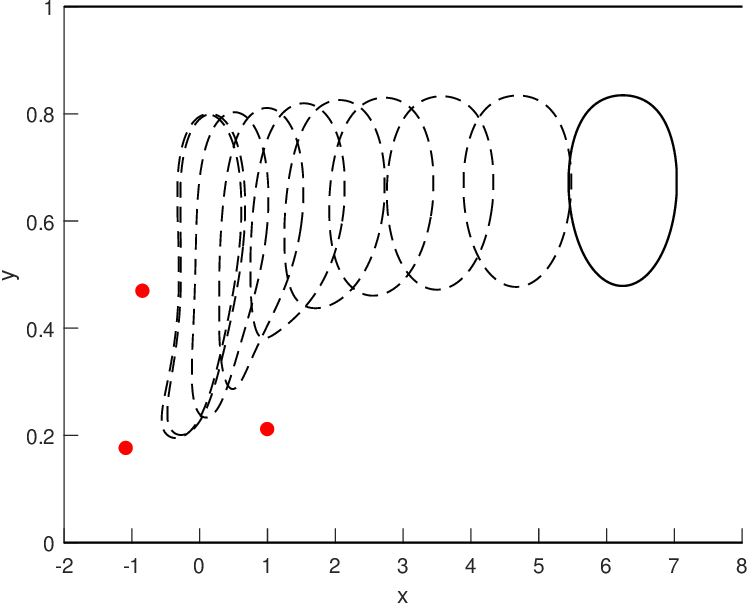}}\qquad
		 \subfigure[\label{fig:2b}]{\includegraphics[width=0.4\textwidth]{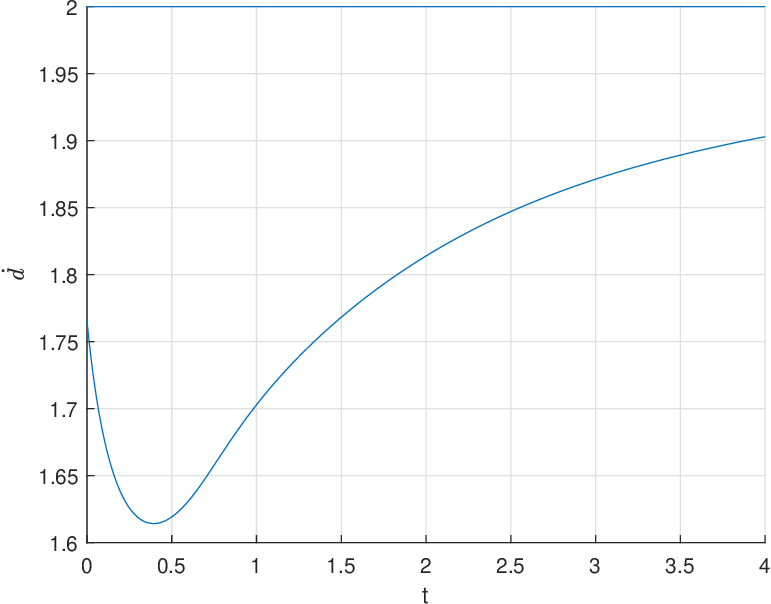}}
		\caption{(a) Bubble evolution leading to a non-symmetric steady shape with  $U=2$. Here the parameters are $\alpha_0=0$,   $\alpha_1=-0.4$, $\alpha_2=\alpha_3=0.2$, $a_1(0)=-0.3$,  $\gamma(0)=\pi$, $a_2(0)=-0.2-0.1i$, $a_3(0)=0.2-0.1i$, and $\rho(0)=0.4$. The bubble shapes are shown for the times $t=0, 0.0343, 0.2373, 0.5241, 0.8454, 1.1490, 1.5136, 1.9591, 2.5337, 3.3132$. (b) The time derivative of the parameter ${d}(t)$ plotted as a function of time.}
			\label{fig:2}
\end{figure}

 In Fig.~\ref{fig:2} we show an example of a bubble evolution where the initial shape is non-symmetric, leading therefore to an asymmetric steady shape. Here the parameters are $\gamma(0)=\pi$, $\alpha_0=0$,  $\alpha_1=-0.4$, $\alpha_2=\alpha_3=0.2$,  $a_1(0)=-0.3$, $a_2(0)=-0.2-0.1i$, and $a_3(0)=0.2-0.1i$.
 It is interesting to see again the strong effect of the singularities of ${\cal S}(z,t)$ on the bubble dynamics:  as the bubble approaches the singularity of ${\cal S}(z,t)$ that lies upstream, it slows down (see Fig.~\ref{fig:2b}) and deforms considerably (see Fig.~\ref{fig:2a}), so as to `overcome' this singularity; and once it does so, the bubble speeds up towards the steady value $U=2$.
 
\begin{figure}[t]
	\center \subfigure[\label{fig:b4a}]{\includegraphics[width=0.4\textwidth]{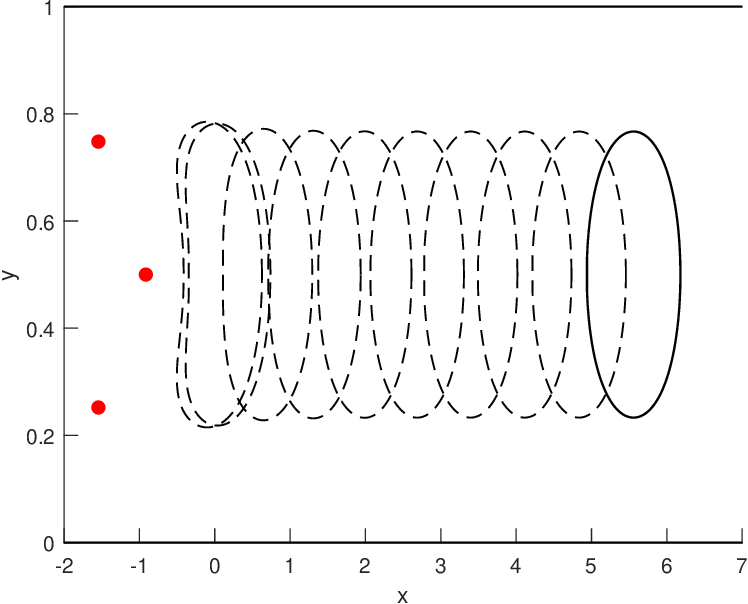}}\qquad\qquad 
		 \subfigure[\label{fig:b4b}]{\includegraphics[width=0.32\textwidth]{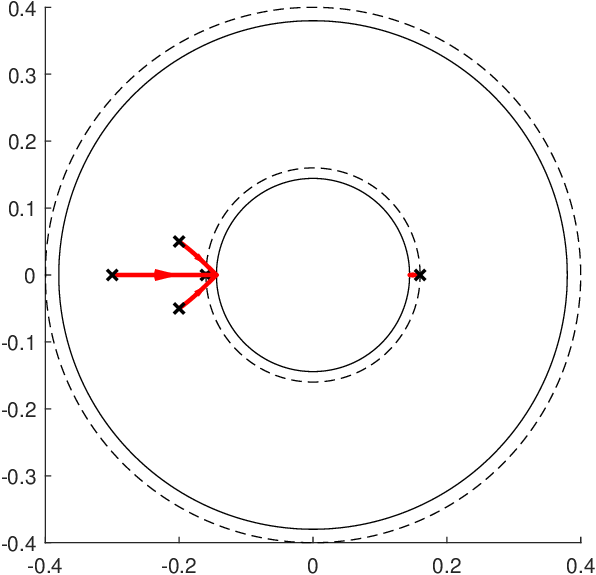}}
		\caption{(a) Bubble evolution  for an initial shape leading to an asymptotic  velocity $U=1.5$. The parameters are (a): $\alpha_0=-0.33$,  $\alpha_1=0.2$, $\alpha_2=-0.4$, $\alpha_3=0.2$, $a_1(0)=-0.2-0.05i$, $a_2(0)=-0.3$, $a_3(0)=-0.2+0.05i$. 
		(b) Trajectories of the singularities of the map $z(\zeta,t)$  in the $\zeta$ plane. }
	\label{fig:b4}
\end{figure}

\begin{figure}[t]
	\center \subfigure[\label{fig:b5a}]{\includegraphics[width=0.4\textwidth]{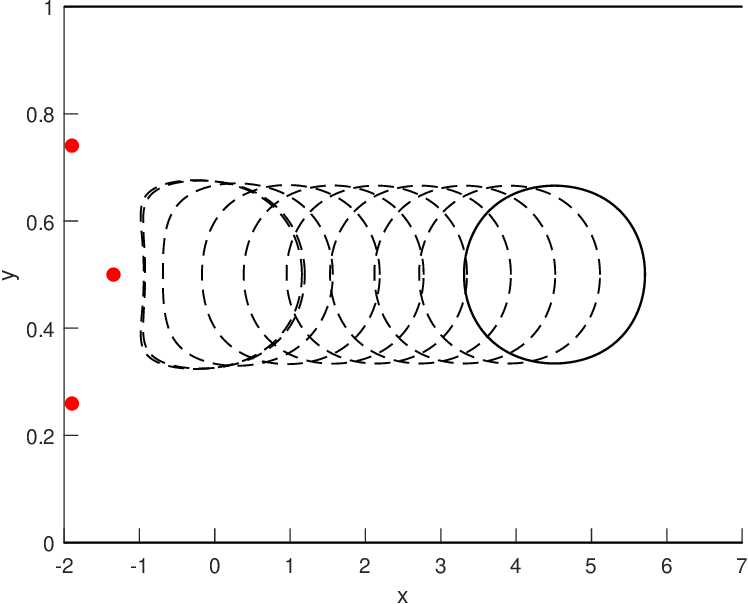}}
		\caption{Bubble evolution leading to $U=2.5$. Here the parameters are: $\alpha_0=0.2$, $\alpha_1=0.2$, $\alpha_2=-0.4$, $\alpha_3=0.2$, $a_1(0)=-0.2-0.05i$, $a_2(0)=-0.3$, $a_3(0)=-0.2+0.05i$.
	 }
	\label{fig:b5}
\end{figure}

Solutions leading to an asymptotic velocity $U\ne2$ are obtained by choosing $\alpha_0\ne 0$. In Fig.~\ref{fig:b4a} we show a case where the final speed is $U=1.5$, with  Fig.~\ref{fig:b4b} showing the trajectories of the singularity of the map $z(\zeta,t)$ in the $\zeta$ plane. Here the parameters are: $\gamma=\pi$, $\alpha_0=-0.33$, $\alpha_1=\alpha_3=0.2$, $\alpha_2=-0.4$, $a_1(0)=\overline{a_3(0)}=-0.2-0.05i$, and $a_2(0)=-0.3$. Notice that  $z(\zeta,t)$ now has singularities at the special points $a_0^+=\rho^ 2$ and $a_0^-=-\rho^ 2$, see (\ref{eq:z0}); besides the singularities $a_k$, $k=1,2,3$, which all approach $a_0^-$ for $t\to\infty$; see Fig.~\ref{fig:b4b}. In Fig.~\ref{fig:b5} we show another example with $U\ne2$, where now the bubble reaches the asymptotic velocity $U=2.5$. Note that bubbles that move with velocity $U<2$ (see Fig.~\ref{fig:b4}) become compressed in the flow direction and elongated in the transverse direction, whereas for bubbles with $U>2$ (see Fig.~\ref{fig:b5}) the opposite happens, as expected \cite{GLV2001}.

\begin{figure}[t]
	\center \subfigure[\label{fig:b3a}]{\includegraphics[width=0.4\textwidth]{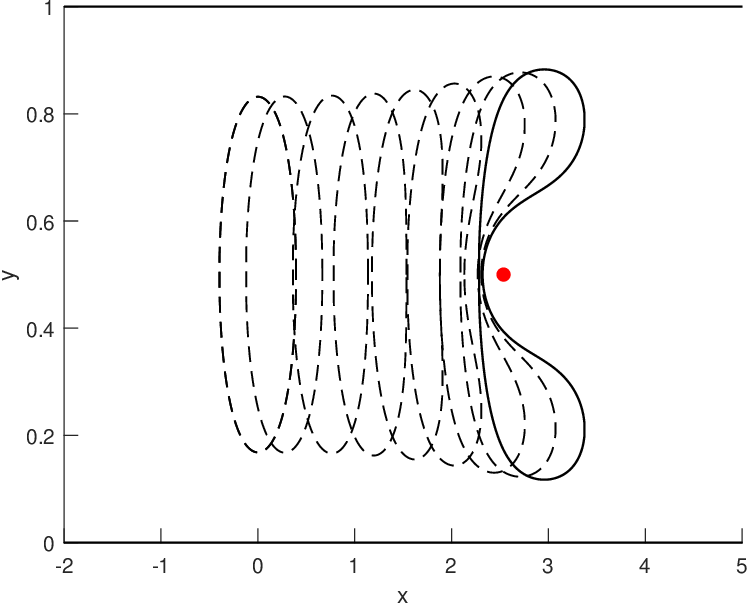}}\qquad
		 \subfigure[\label{fig:b3b}]{\includegraphics[width=0.4\textwidth]{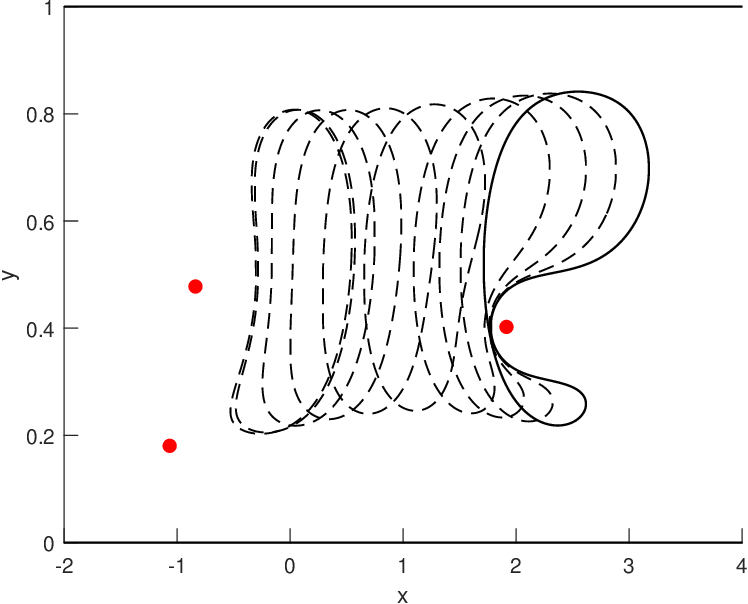}} 
   \caption{Time evolution of  symmetric (a)  and  non-symmetric  (b)  bubbles, showing   `breakup' at finite time. The parameters are (a): $\alpha_0=0$,  $\alpha_1=-0.3$, $\alpha_2=0.3$, $a_1(0)=-0.17$, $a_2(0)=0.18$; and (b): $\alpha_0=0$,  $\alpha_1=0.2$, $\alpha_2=-0.4$, $\alpha_3=0.2$, $a_1(0)=-0.2-0.1i$, $a_2(0)=-0.3$, $a_3(0)=0.2-0.015i$ }
	\label{fig:b3}
\end{figure}

\begin{figure}[t]
	\center 
   \subfigure[\label{fig:b3c}]{\includegraphics[width=0.4\textwidth]{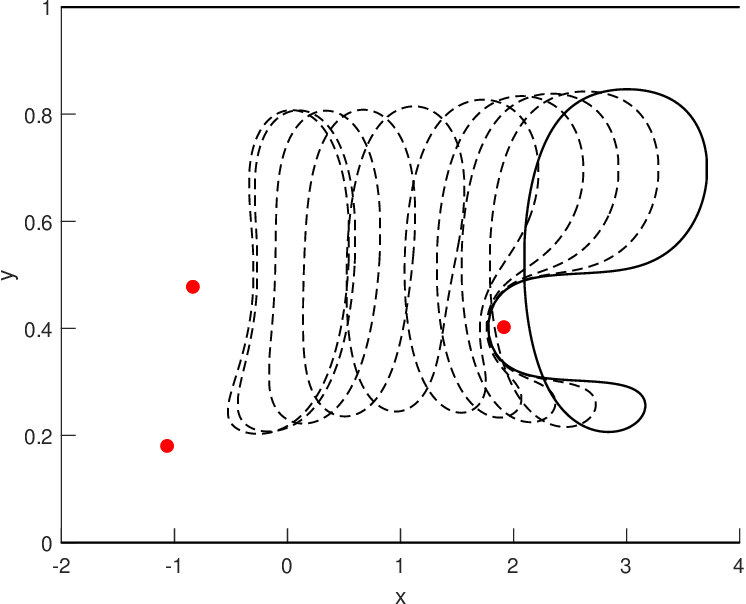}}\qquad
		 \subfigure[\label{fig:b3d}]{\includegraphics[width=0.3\textwidth]{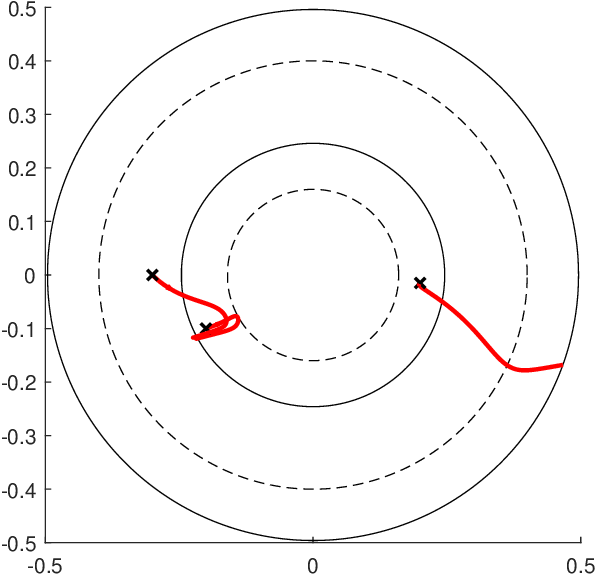}}
   \caption{Mathematical evolution in the physical $z$ plane (a) and mathematical $\zeta$ plane (b) of  a logarithmic solution past the time of the initial loss of univalence of the map $z(\zeta,t)$. The parameters are as in Fig.~\ref{fig:b3b}.}
	\label{fig:b3cross}
\end{figure}

If the initial conditions are outside the set of well-posedness, it may so happen that the bubble does not succeed to get past the singularities of ${\cal S}(z,t)$, leading to a loss of univalence of the map $z(\zeta,t)$ at some finite time, at which point the interface self-intersects and the solution ceases to be physically valid henceforth.  Such bubble `breakup' will always occur, for example, if we start with a symmetric shape having a singularity of ${\cal S}(z,t)$  on the channel centreline  ahead of the bubble, an example of which is shown in Fig.~\ref{fig:b3a}. One sees from this figure that, as the bubble approaches the singularity (red dot) of  ${\cal S}(z,t)$, the fore side of the interface gets `pinned' by the singularity, so that an incipient `fjord' develops. However, as the aft-side of the interface continues to advance, it will eventually touch the fore interface. (The last interface shown in Fig.~\ref{fig:b3a} corresponds to shortly before this touching point.) Asymmetric bubble breakup can also occur,  as shown in Fig.~\ref{fig:b3b}. In this figure, the initial shape is  such that the singularity of ${\cal S}(z,t)$ that induces the breakup is closer to the bottom part of the interface. In real systems, once a very narrow bubble neck forms, other physical effects come into play causing the bubble to  split into two smaller bubbles, which would then continue to move down the channel, possibly reaching a steady  multibubble configuration  \cite{PhysicaD2023}. Exact solutions such as those shown in Fig.~\ref{fig:b3} can thus be viewed as  describing the dynamical process leading to bubble breakup. Although these solutions are not physically valid past the time when the interface first self-intersects, they remain nonetheless mathematical valid in the sense that the corresponding dynamical system for the singularities $\{a_k(t)\}$ can be integrated past the `breakup' point, as shown in Fig~\ref{fig:b3cross}.


\section{Rational Solutions}
\label{sec:rational}

Here we consider solutions where the function $h(\zeta,t)$ has only simple poles. (Solutions with poles of higher order could in principle be constructed but we shall not pursue this detail here.) More concretely, we write
 \begin{align}
h(\zeta, t) =\sum_{k=1}^N  \left[\alpha_k   K(a_k/\zeta;\rho^2)  + \overline{ \alpha}_k K(\overline a_k \zeta;\rho^2)\right],
\label{eq:hr}
\end{align}
where  $\alpha_k(t)$ and $a_k(t) \notin D_\zeta$, $k=1,..,N$, are complex time-dependent parameters, and the function $K(\zeta;\rho)$ is defined as
\begin{align}
K(\zeta;\rho)=\zeta\frac{d}{d\zeta} \log P(\zeta;\rho),
\end{align}
or more explicitly
\begin{align}
K(\zeta;\rho)=\frac{1}{\zeta^{-1}-1} + \sum_{n=0}^\infty \rho^{2n}\left[\frac{1}{\zeta-\rho^{2n}}-\frac{1}{\zeta^{-1}-\rho^{2n}}\right].
\label{eq:K}
\end{align}
It is easy to very the following properties of the function $K(\zeta;\rho)$:
\begin{align}
K(1/\zeta;\rho)&=1-K(\zeta;\rho),
\label{eq:PK1}
\\
K(\rho^{\pm2}\zeta;\rho)&=K(\zeta;\rho) \mp 1,
\label{eq:PK2}
\end{align}
which in turn imply that the function $h(\zeta,t)$ given in (\ref{eq:hr}) satisfies condition (\ref{eq:barh}) and (\ref{eq:hrho}), as required.

\begin{figure}[t]
	\center \subfigure[\label{fig:case4a}]{\includegraphics[width=0.45\textwidth]{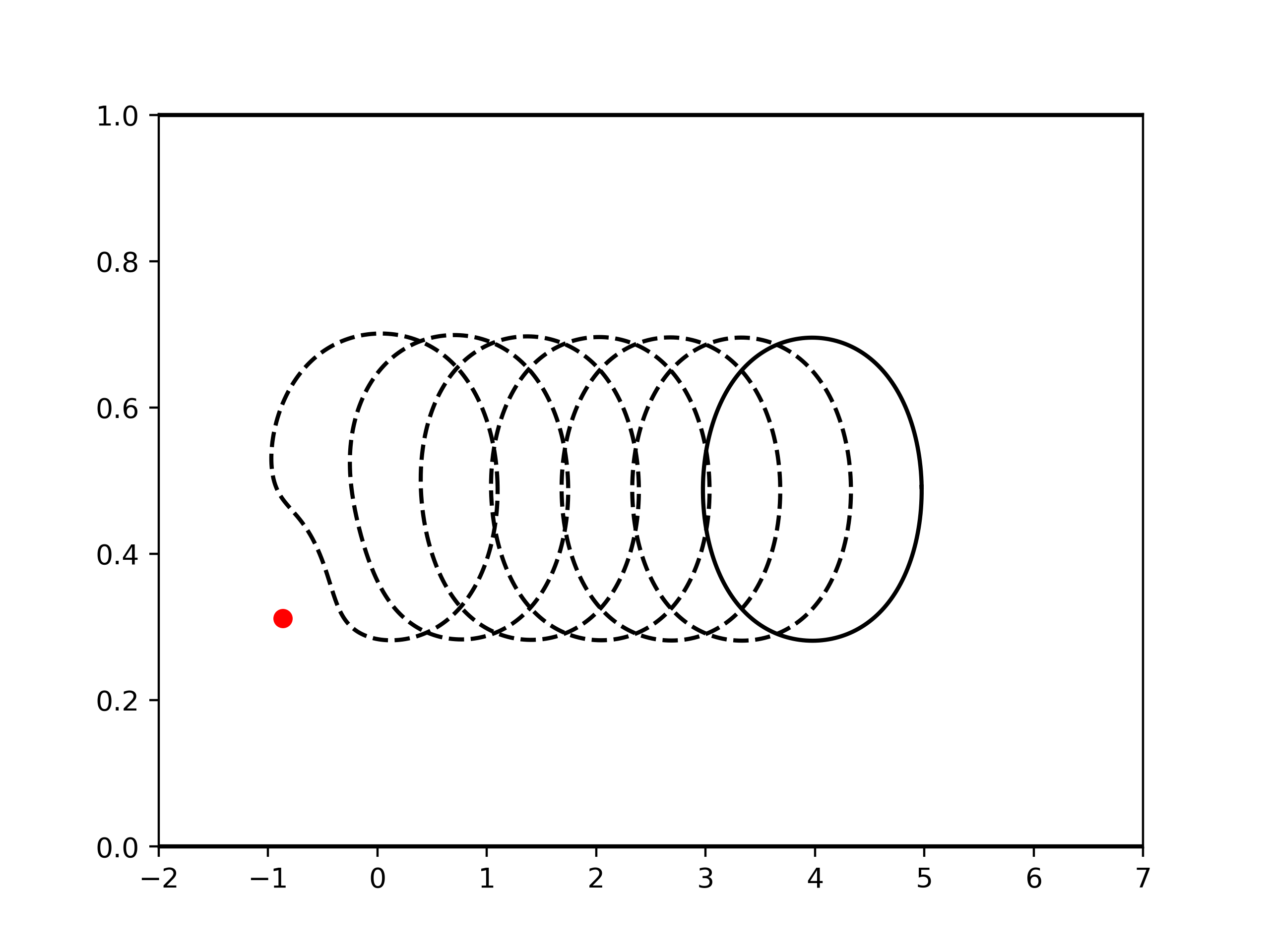}}\qquad
		 \subfigure[\label{fig:case4b}]{\includegraphics[width=0.41
		  \textwidth]{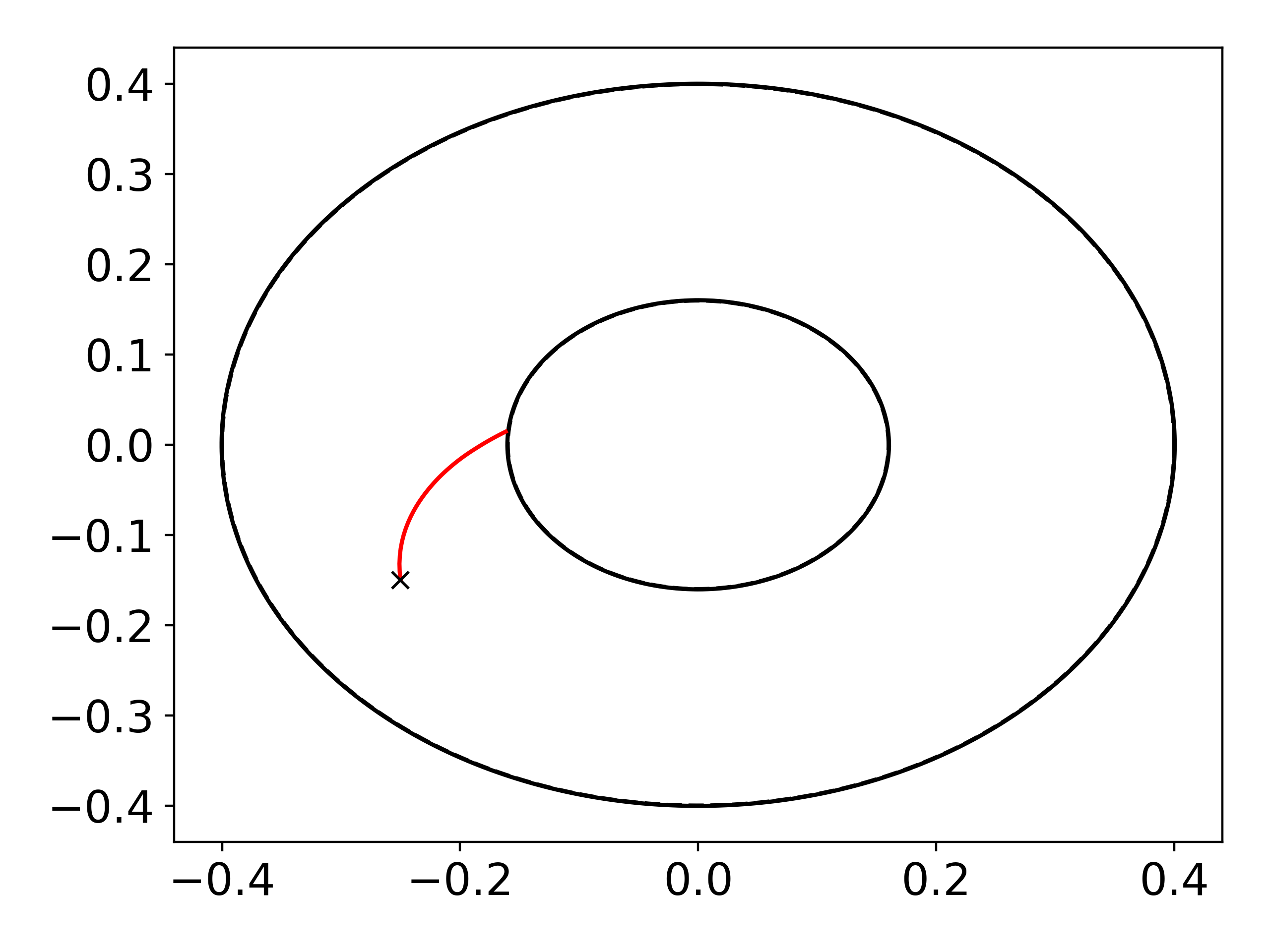}}
		 \subfigure[\label{fig:case4c}]{\includegraphics[width=0.45\textwidth]{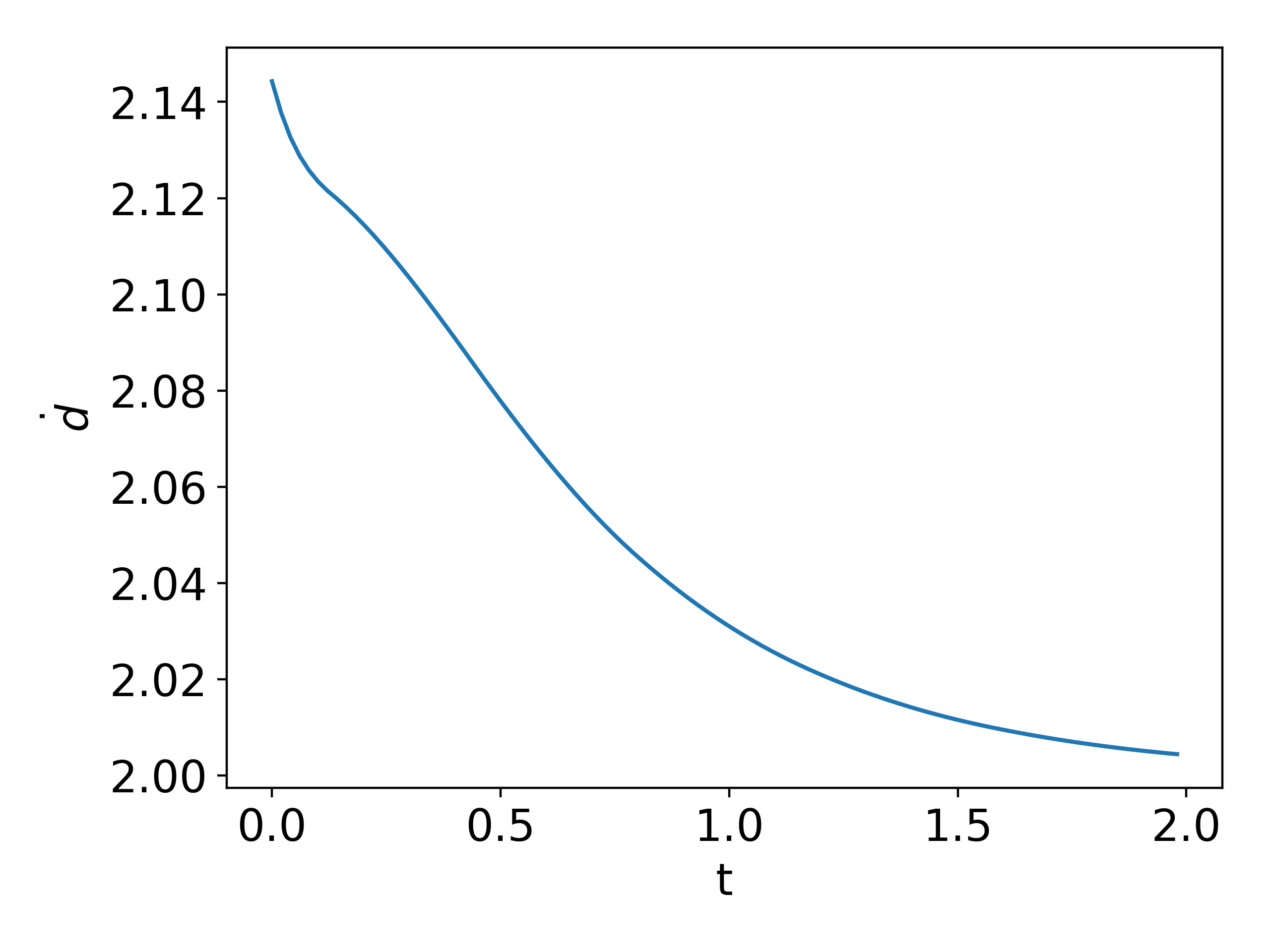}}
		\caption{(a) Bubble evolution leading to a non-symmetric steady shape with  $U=2$ for a rational solution with one pole singularity. Here the parameters are $\alpha_0=0$,   $\alpha_1=-0.05-i0.05$    , $a_1(0)=-0.55-i0.15$, $\gamma(0)=\pi$, and $\rho(0)=0.16$. 
The bubble shapes are shown at equal time intervals from $t=0$ up to $t=2$. (b) Time evolution of the singularity $a_1(t)$ in the complex $\zeta$ plane. (c) Time derivative of the parameter ${d}(t)$ plotted as a function of time.}
\label{fig:case4}
\end{figure}

\begin{figure}[t]
	\center \subfigure[\label{fig:case7a}]{\includegraphics[width=0.45\textwidth]{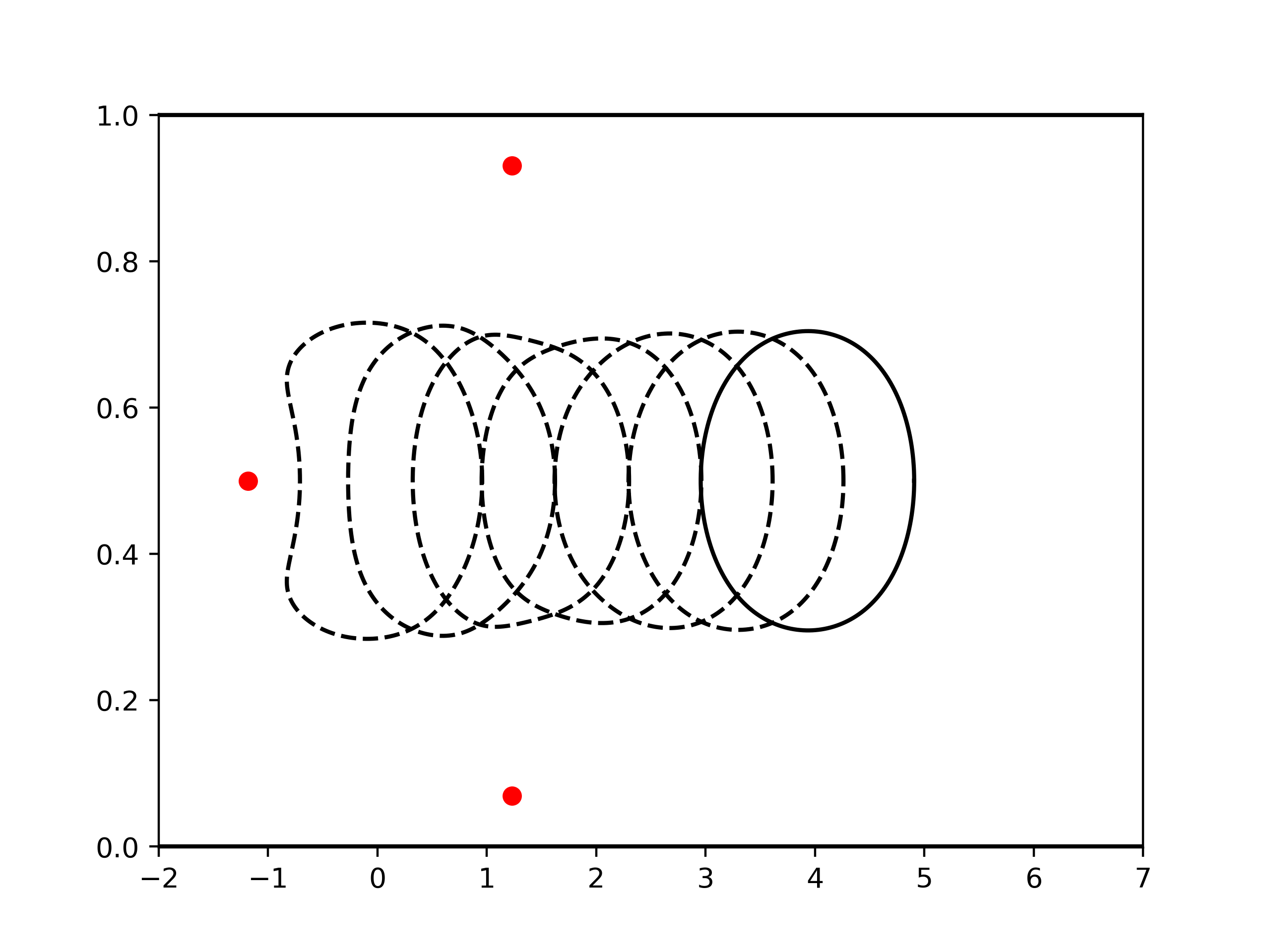}}\qquad
		 \subfigure[\label{fig:case7b}]{\includegraphics[width=0.41
		  \textwidth]{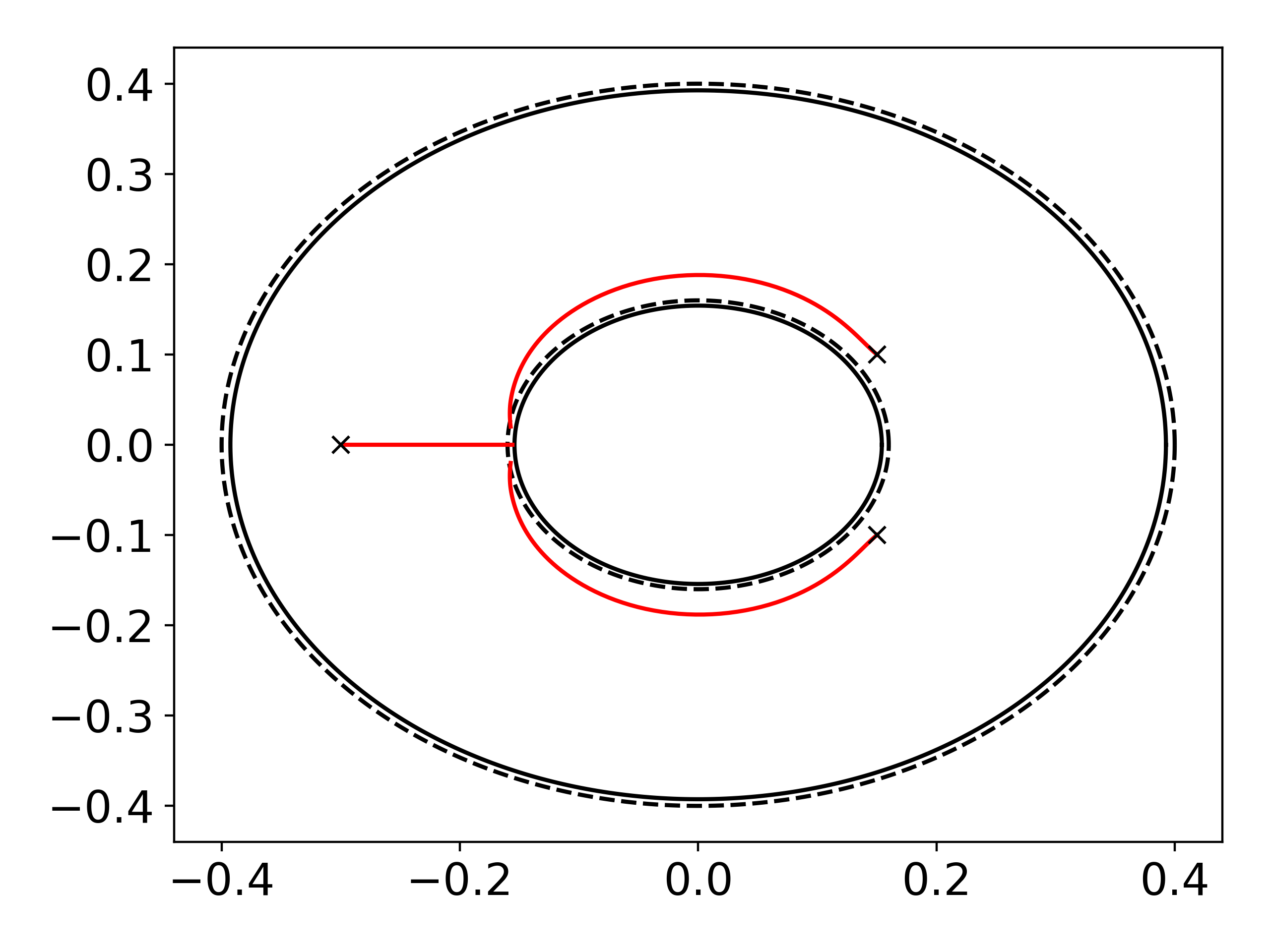}}
		 \subfigure[\label{fig:case7c}]{\includegraphics[width=0.45\textwidth]{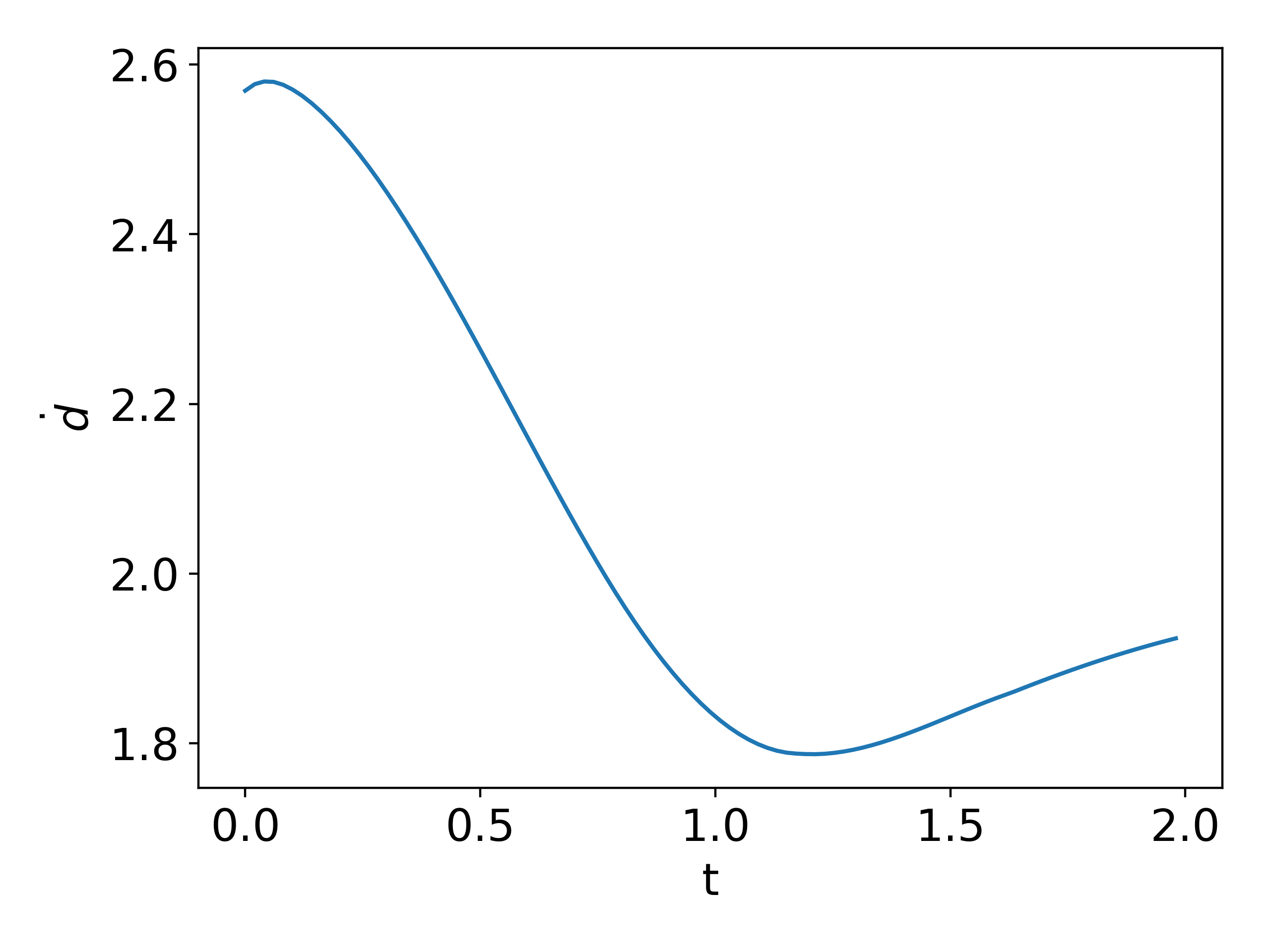}}
		\caption{(a) Bubble evolution leading to a symmetric steady shape with  $U=2$ for a rational solution with three pole singularities. Here the parameters are $\alpha_0=0$,   $\alpha_1={\alpha}_3=0.05$, $\alpha_2=-0.02$, $a_1(0)=\overline{a}_3=0.15+i0.1$, $a_2(0)=-0.1$, $\gamma(0)=\pi$
		and $\rho(0)=0.16$. The bubble shapes are shown at equal time intervals from $t=0$ up to $t=2$. (b) Time evolution of the singularity $a_k(t)$, $k=1,2,3$, in the complex $\zeta$ plane. (c) Time derivative of the parameter ${d}(t)$ plotted as a function of time.}
			\label{fig:case7}
\end{figure}

\begin{figure}[t]
	\center \subfigure[\label{fig:case9a}]{\includegraphics[width=0.45\textwidth]{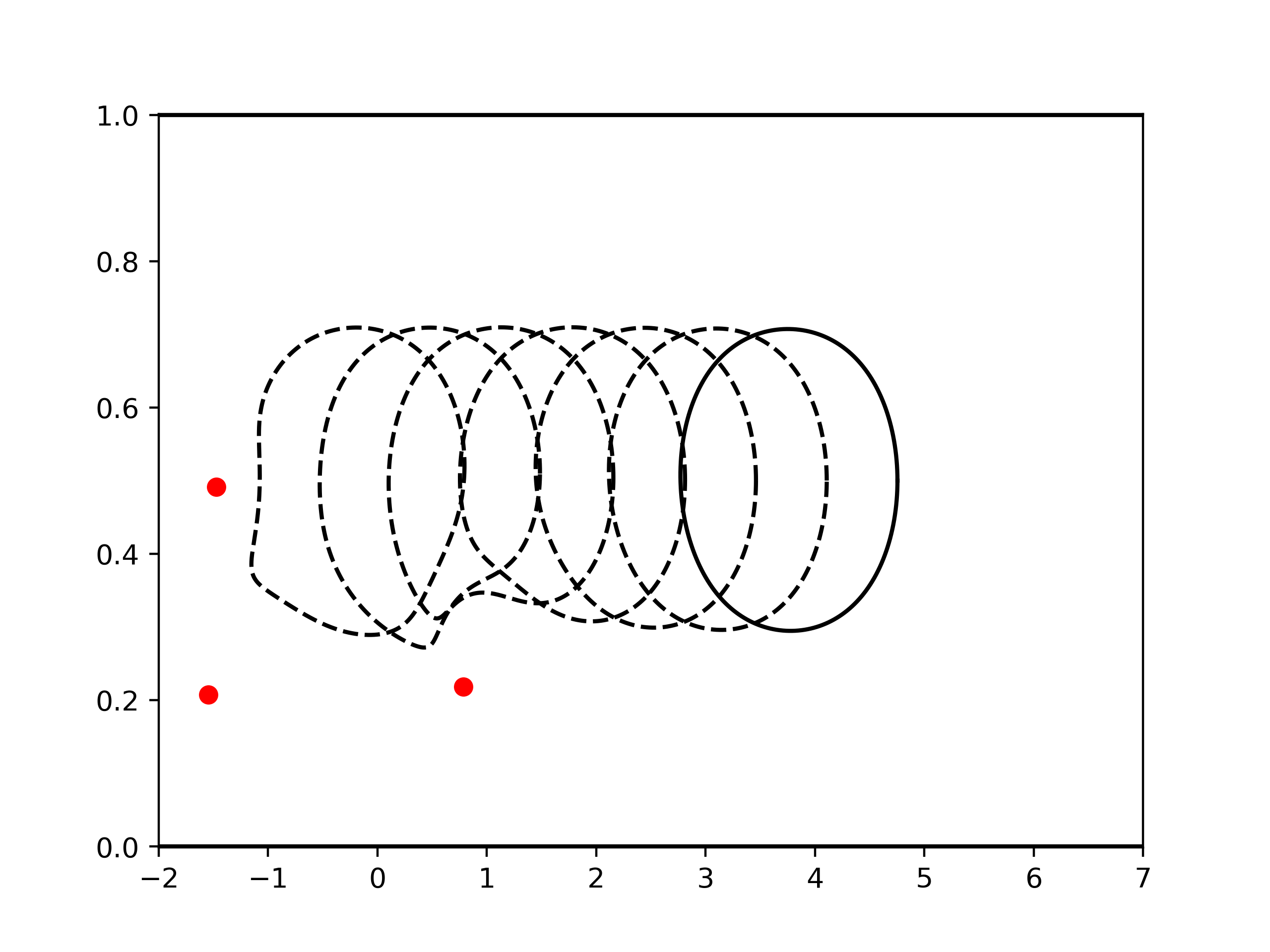}}\qquad
		 \subfigure[\label{fig:case9b}]{\includegraphics[width=0.41
		  \textwidth]{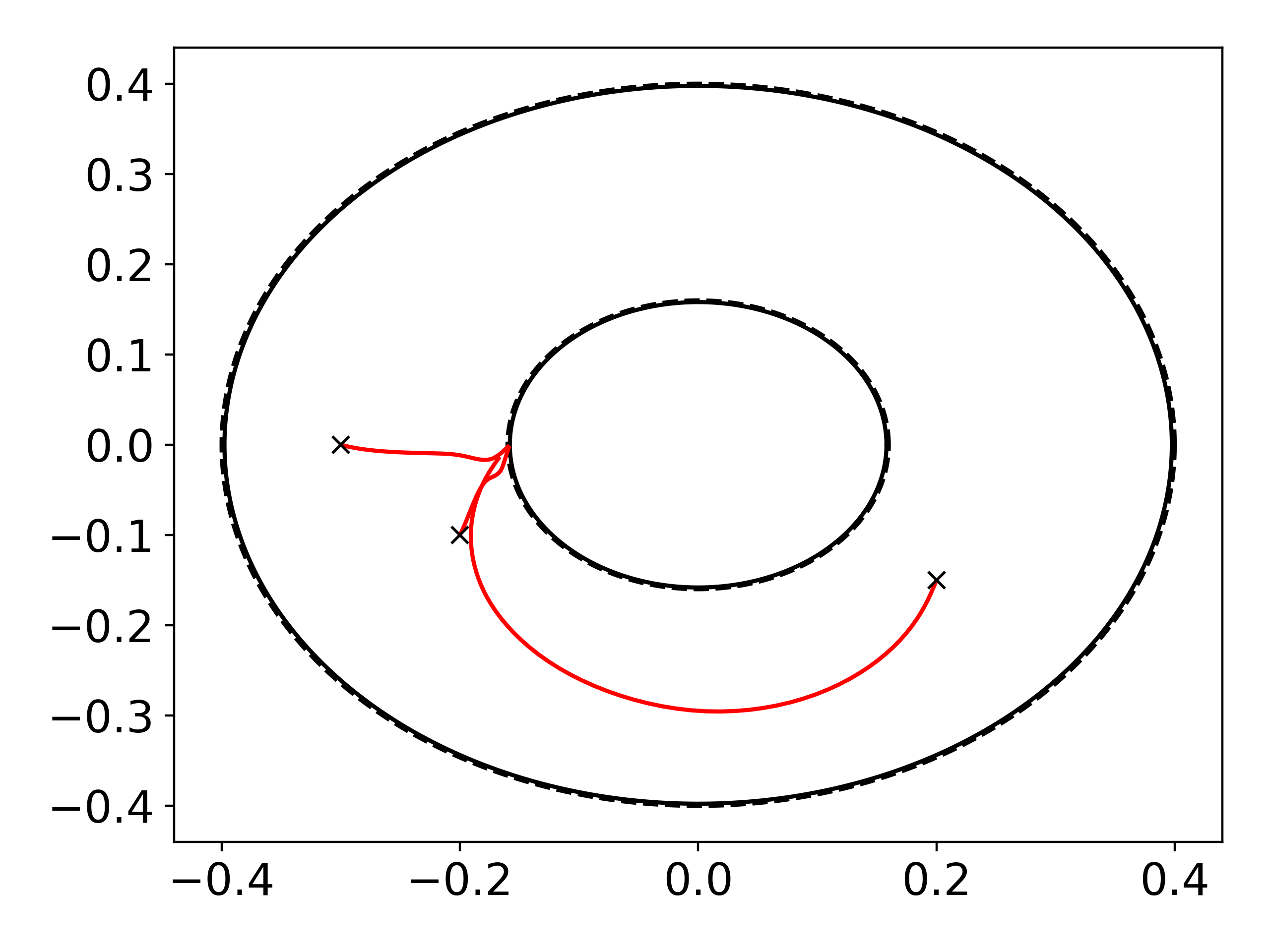}}
		 \subfigure[\label{fig:case9c}]{\includegraphics[width=0.45\textwidth]{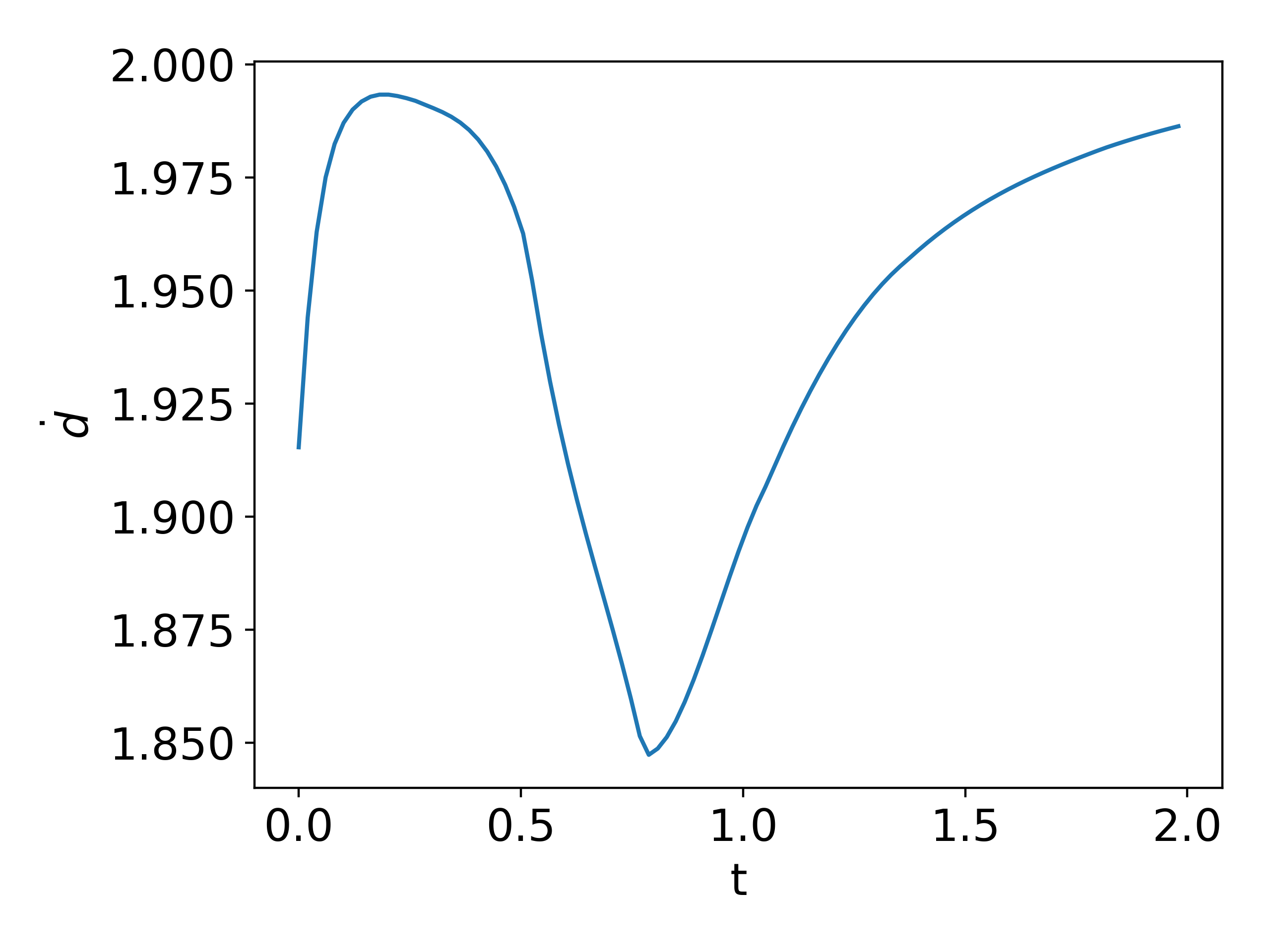}}
		\caption{(a) Bubble evolution leading to a non-symmetric steady shape with  $U=2$ for a rational solution with three pole singularities. Here the parameters are $\alpha_0=0$,   $\alpha_1=0.1$, ${\alpha}_2=-0.05$, $\alpha_3=0.05$, $a_1(0)=-0.2-i0.1$, $a_2(0)=-0.3$, ${a}_3=0.2-i0.15$,
$\gamma(0)=\pi$, and $\rho(0)=0.6$. 
		The bubble shapes are shown at equal time intervals from $t=0$ up to $t=2$. (b) Time evolution of the singularity $a_k(t)$, $k=1,2,3$, in the complex $\zeta$ plane. (c) Time derivative of the parameter ${d}(t)$ plotted as a function of time.}
			\label{fig:case9}
\end{figure}

The conserved quantities $\beta_k$ and $\beta_\pm$ in this case are  given by
 \begin{align}
\beta_k = & d(t) +i\frac{\gamma}{2}+ \log \frac{P(e^{-i\gamma}\rho^2/\overline a_k;\rho^2)}{P(\rho^2/\overline a_k;\rho^2)}  +   \alpha_0 \log\frac{P(e^{i\gamma}\overline a_k; \rho^2)}{P(\overline a_k;\rho^2)}
 + \cr
 +& \sum_{m=1}^N \left[ \alpha_m K(a_m \overline a_k /\rho^2;\rho^2) + \overline \alpha_m K(\rho^2 \overline a_m /\overline a_k;\rho^2)\right],
\label{eq:betakr}
\end{align}
and 
\begin{align}
\beta_\pm = &   (1-\alpha_0) d(t) -2t \pm
(1 -\alpha_0^2) \log \frac{P(\rho^2 e^{i\gamma};\rho^2)}{P(\rho^2;\rho^2)} - i\frac{\gamma}{2}\left(1+\alpha_0\right) +\cr 
&+ \sum_{k=1}^N \left[ \alpha_k K(\rho^2a_k/\zeta_\pm;\rho^2) +  \overline \alpha_k K(\overline a_k \zeta_\pm/\rho^2 ;\rho^2) \right]\cr
&- \alpha_0 \sum_{k=1}^N \left[ \alpha_k K(a_k/\zeta_\pm;\rho^2) + \overline \alpha_k K(\overline a_k \zeta_\pm;\rho^2)\right].
 \label{eq:betapmr}
\end{align}
From (\ref{eq:hr}) one also  sees that the residue of $h(\zeta,t)$ at a pole $a_k$ is $r_k=\alpha_k a_k$; hence the additional conserved quantity $R_k$ introduced in (\ref{eq:Rk}) becomes
\begin{align}
 R_k = -\frac{\rho^2\overline \alpha_k}{\overline a_k}z_\zeta(\rho^2/\overline a_k).
 \label{eq:Rkr}
 \end{align}
The set of conserved quantities, $\beta_\pm$ and $\{\beta_k, R_k | k=1,...,N\}$, are thus sufficient to determine the time evolution of the parameters $d(t)$, $\gamma(t)$, $\rho(t)$, and $\{a_k, \alpha_k | k=1,...,N\}$. Examples of solutions for this case are shown in Figs.~\ref{fig:case4}-\ref{fig:case9}. In the examples shown we have chosen $\alpha_0=0$, so that the bubble will reach a steady shape with  $U=2$.

In Fig.~\ref{fig:case4a} we show a rational solution that has only one simple pole, i.e., $N=1$ in (\ref{eq:hr}). As before, the red point in this figure indicate the location of the fixed singularity of the Schwarz function.  Here the initial values of the parameters are  $\alpha_1(0)=-0.05-i0.05$, $a_1(0)=-0.55-i0.15$, $\gamma(0)=\pi$, and $\rho(0)=0.16$. The bubble shapes are shown at equal time intervals from $t=0$ up to $t=2$. In Fig.~\ref{fig:case4b} we show the time evolution of the singularity $a_1(t)$ in the complex $\zeta$ plane, where one sees that it approaches the attractor $a_0^-=\rho^2 e^{i\gamma}$, and as it does so the bubble  reaches a steady shape with $U=2$, as indicated in Fig.~\ref{fig:case4c}, where  the time derivative of the parameter ${d}(t)$ is plotted as a function of time.
Similar behavior is seen in Fig.~\ref{fig:case7}, where we show a symmetric rational solution with three poles ($N=3$), with the following initial parameters: $\alpha_1(0)={\alpha}_3(0)=0.05$, $\alpha_2(0)=-0.02$, $a_1(0)=\overline{a}_3(0)=0.15+i0.1$, $a_2(0)=-0.1$, $\gamma(0)=\pi$, and $\rho(0)=0.16$.

\begin{figure}[t]
	\center \subfigure[\label{fig:cusp_ra}]{\includegraphics[width=0.49\textwidth]{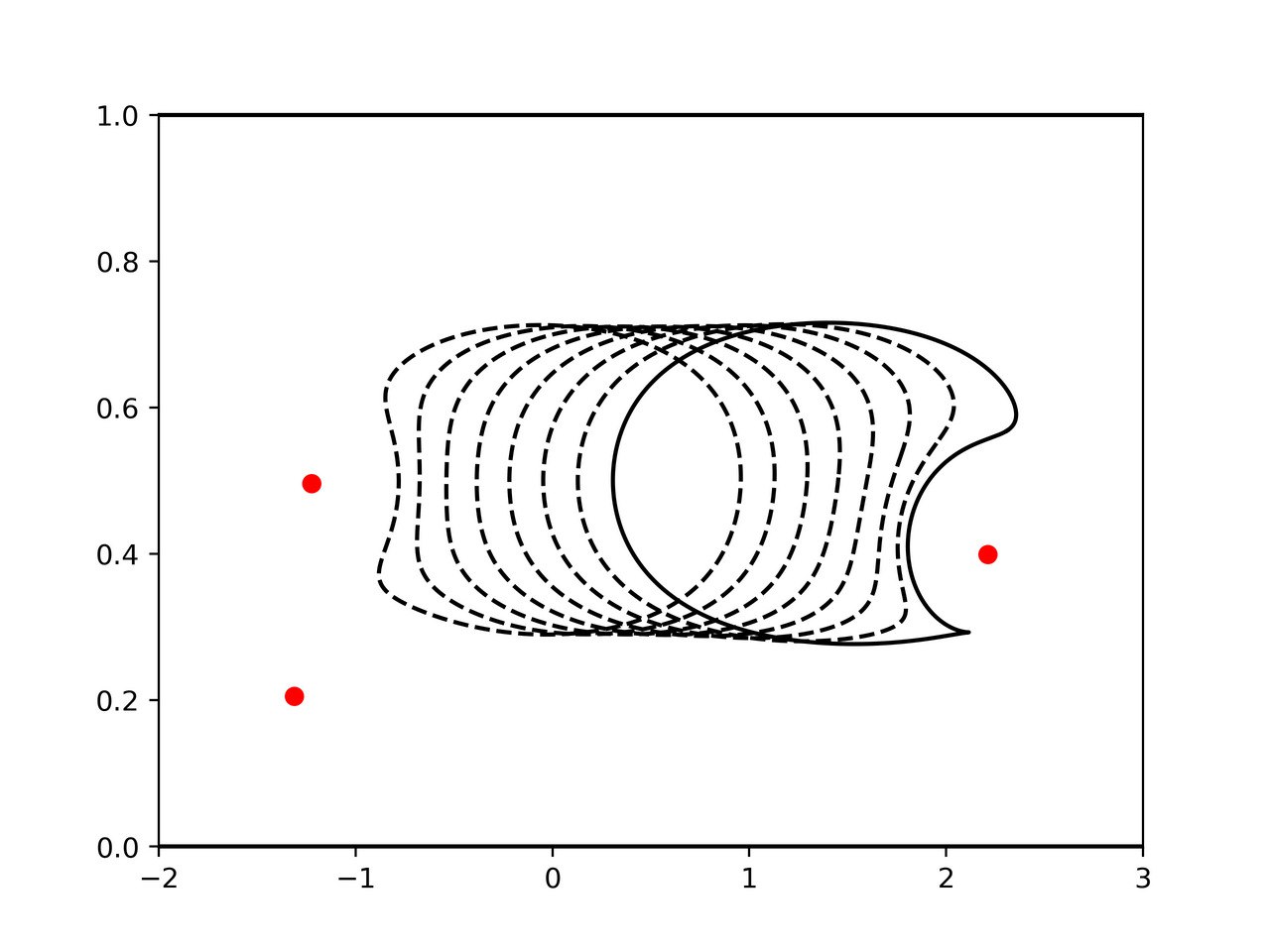}}\quad
		\subfigure[\label{fig:cusp_rb}]{\includegraphics[width=0.45\textwidth]{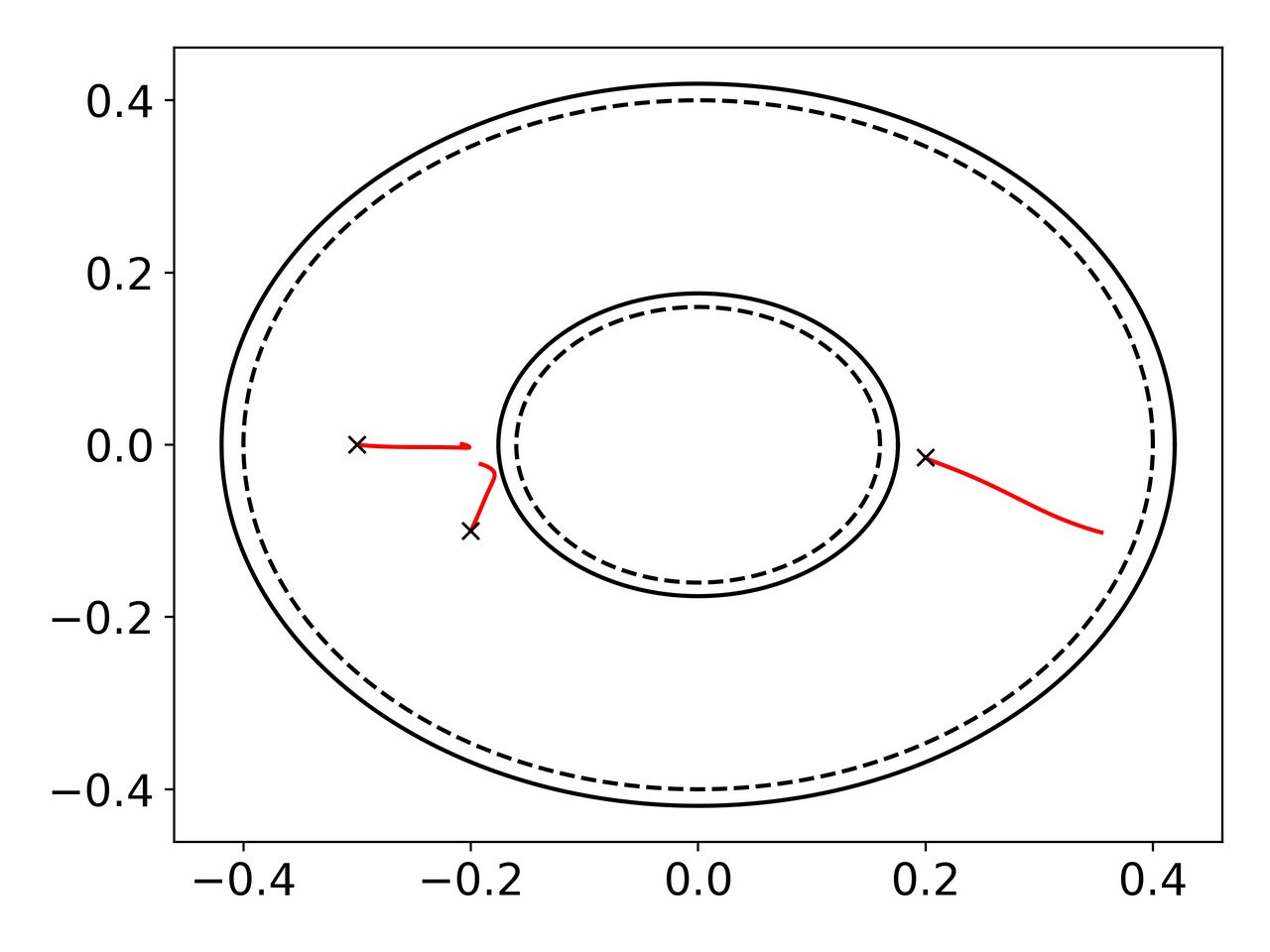}}
		\caption{(a) Bubble evolution leading to cusp formation in finite time. Here the parameters are $\alpha_0=0$,   $\alpha_1(0) = 0.04$, $\alpha_2(0) = -0.08$, $\alpha_3(0) = 0.04$, $a_1(0) = -0.2 - 0.1i$, $a_2(0) = -0.3$, $a_3(0) = 0.2 - 0.015i$,
$\gamma(0)=\pi$, and $\rho(0)=0.16$. 
		The bubble shapes are shown at equal time intervals of $t=0$ up to $t=0.65$. 
  (b) Time evolution of the singularity $a_k(t)$, $k=1,2,3$, in the complex $\zeta$ plane. 
  }
			\label{fig:cusp_r}
\end{figure}

\begin{figure}[h]
	\center \subfigure[\label{fig:rcross_a}]{\includegraphics[width=0.49\textwidth]{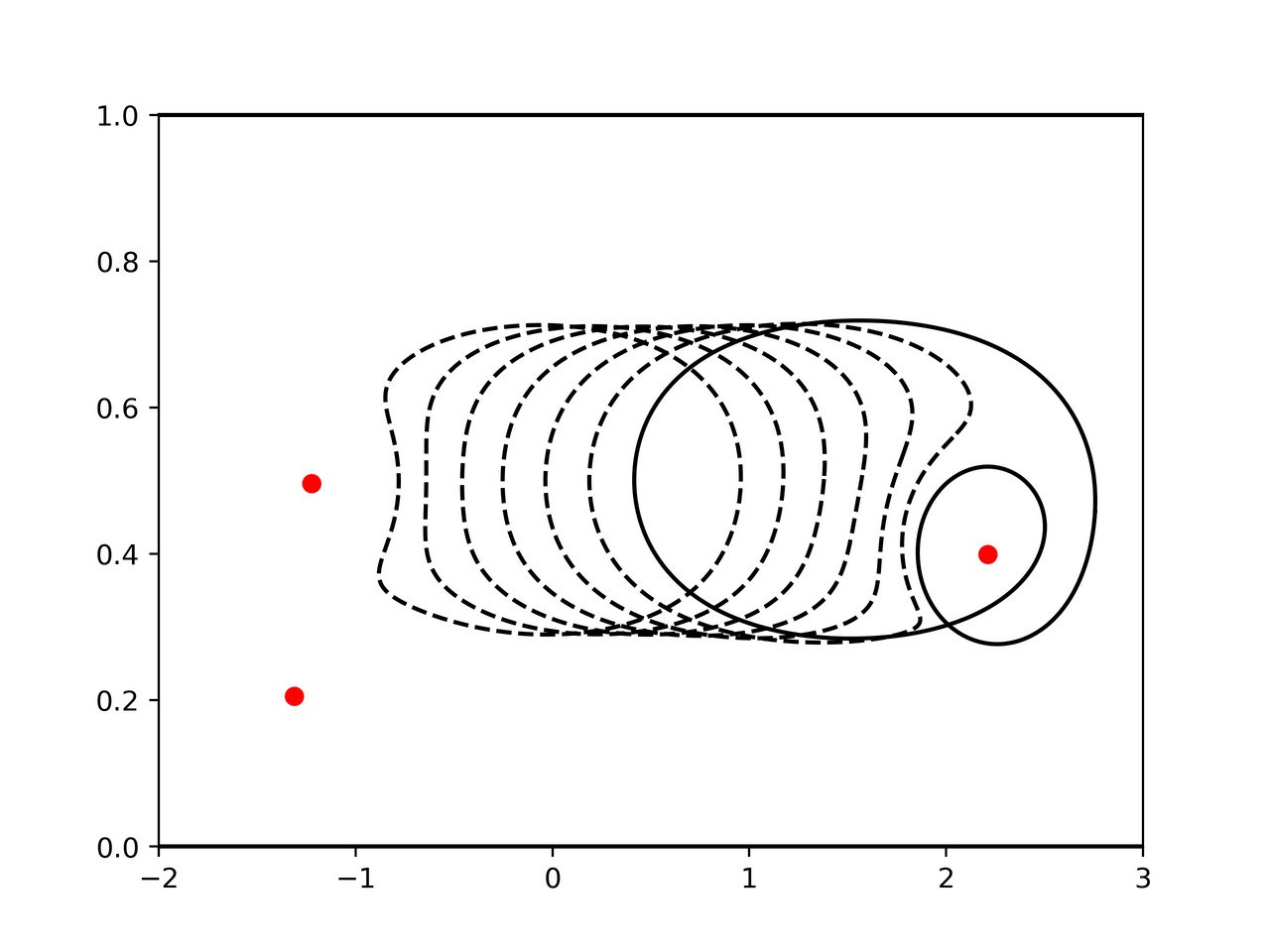}}\quad
		\subfigure[\label{fig:rcross_b}]{\includegraphics[width=0.45\textwidth]{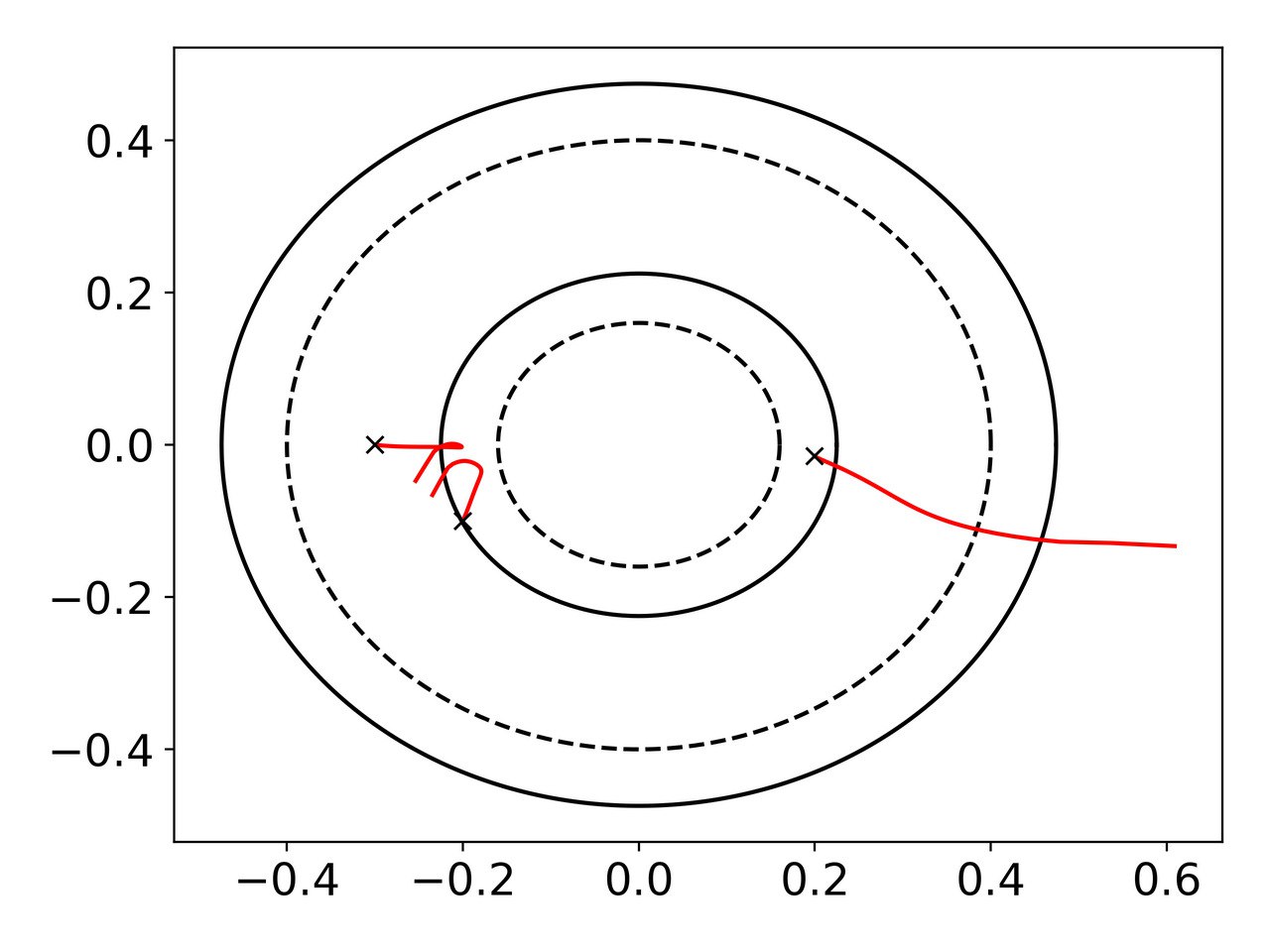}}

  \caption{Mathematical evolution in the physical $z$ plane (a) and mathematical $\zeta$ plane (b) of  a rational solution past the time of cusp formation. The parameters are as in Fig.~\ref{fig:cusp_r}.}
			\label{fig:rcross}
\end{figure}

Depending on the location of the singularities of the Schwarz function of the initial shape (which we recall remain fixed), the bubble may develop ``sharp corners'' as the interface approaches one of these fixed points. As example of this situation is shown in Fig.~\ref{fig:case9}, where one sees that the interface deforms quite a bit and develops a region of high curvature (``near cusp'') as it approaches the singularity of ${\cal S}(z,t)$ that originally lies ahead of the bubble, which causes the bubble to slow down momentarily, as seen in Fig.~\ref{fig:case9c}. But once the bubble manages to overcome this singular point (of the Schwarz function),  it quickly approaches a steady-state shape with $U=2$; see Fig.~\ref{fig:case9c}. There may however be initial data that leads to actual cusp formation in finite time, as shown in Fig.~\ref{fig:cusp_r}, where the last interface shown is for the time when the cusp forms, which corresponds to the time when a zero of the mapping $z(\zeta,t)$ hits the circle $C_1$ (of radius $\rho$). After this time the solution is no longer physically acceptable  (because the interface self-intersects) although it remains mathematically valid in the sense that the numerical integration of the corresponding dynamical system is still possible, as shown in Fig.~\ref{fig:rcross}.


\section{Velocity Selection}
\label{sec:selection}

As already mentioned in the introduction, the {\it selection problem} for Hele-Shaw flows
concerns the fact that out of the continuous family of  analytical solutions for a steadily moving finger \cite{ST} or bubble \cite{TS} only a specific pattern  (namely, that with velocity $U=2$) is experimentally observed
(in the limit of vanishing surface tension). In the mid 1980s, it was shown by several groups \cite{SurfaceTension} that the inclusion of small surface tension effects can select the observed pattern. 
Subsequently, it was shown \cite{98PRL} from the analysis of time-dependent solutions without surface tension which remain non-singular for all times, that the selected finger, i.e., that with $U=2$, is the only attractor of the dynamics. This result indicated for the first time that surface tension was not necessary for velocity selection, as the special nature of the selected pattern is already present in the dynamics of the idealized problem (i.e., without regularizing boundary conditions). 
Recently, it was shown that a similar selection mechanism also holds for Hele-Shaw bubbles 
both in a channel \cite{us2014,PhysicaD2023} and in an unbounded cell \cite{Robb2015}.  
In what follows we give a more complete presentation of the argument, briefly discussed in Ref.~\cite{us2014}, that addresses the selection problem for a single bubble in a Hele-Shaw channel through the stability analysis of the time-dependent solutions described in Sec.~\ref{sec:exact}.

First, we recall, as shown in Sec.~\ref{sec:asymptotic}, that if assume that the function $z(\zeta,t)$ given in (\ref{eq:zg}) exists for all times, then the function $h(\zeta,t)$ must vanish for $t\to\infty$, leaving us with the steady solution (\ref{eq:z0}) where the bubble moves with speed $U$. Now, a special aspect of the steady solutions  with $U\ne 2$ is that the mapping $z(\zeta,t)$ given in (\ref{eq:z0})  have singularities at  the two special points  $a_0^+=\rho^2$ and $a_0^-=\rho^2e^{i\gamma}$, implying that the Schwarz function ${\cal S}(z,t)$ has a pole at infinity. Suppose now that we perturb this steady solution slightly by causing an infinitesimal displacement of these two singularities.  
More precisely, we consider a perturbed initial shape of the form   
\begin{equation}
\begin{split}
\label{eq:zbp}
z(\zeta, 0) =~& d(0)+ i\frac{\gamma}{2}+ \log \frac{P(e^{-i\gamma}\zeta;\rho^2)}{P(\zeta;\rho^2)} 
 +   \frac{\alpha_0}{2}  \log \frac{ P(\zeta/(\rho^2e^{i\gamma}+\varepsilon_-))P(1/\zeta(\rho^2e^{-i\gamma}+\overline\varepsilon_-))} {P(\zeta/(\rho^2 +\varepsilon_+))P(1/\zeta(\rho^2 +\overline\varepsilon_+))}.
 \end{split}
\end{equation}
where $\varepsilon_\pm(0)$ are some small perturbations, i.e., $|\varepsilon_\pm(0)|\ll 1$.
The form of the perturbation term above is dictated by the requirement that it must have constant imaginary part on $C_0$, so as to  preserves the boundary condition ${\rm Im}[z(\zeta,t)]=\mbox{const.}$,  for $\zeta\in C_0$. (The perturbation must  belong to the set of well-posedness, which is always possible to accommodate.)

Note that moving the singularities of the map $z(\zeta,t)$ from  the special points $a_0^\pm$ to arbitrarily nearby  points,  $a_1=a_0^-+\varepsilon_-$ and $a_2=a_0^+ +\varepsilon_+$, has an arbitrarily small effect on the actual shape of the perturbed interface. Indeed, one can  easily verify  that  Eq.~(\ref{eq:zbp})  recovers  the corresponding unperturbed solution (\ref{eq:z0})  for $\varepsilon_\pm\to 0$.  [To show this, use property (\ref{eq:P2}) which implies that $P(1/\zeta\rho^2e^{-i\gamma})/P(1/\zeta\rho^2)=e^{i2\gamma}P(e^{-i\gamma}\zeta/\rho^2)/P(\zeta/\rho^2)$.] This small perturbation has, notwithstanding, a profound impact on the singularity structure of the Schwarz function, in that  it replaces the pole  of  ${\cal S}(z,t)$ at infinity with two finite logarithmic singularities. 
(Furthermore,  the perturbations must be such that $\rho^2 < |a_1|, |a_2| < \rho$, so that $z(\zeta,t)$ remains free from singularities in the fluid domain, except at the preimages, $\zeta_\pm$, of the points $x=\pm\infty$.)

Since the perturbed map (\ref{eq:zbp}) now has singularities at the  points $a_1\ne a_0^-$ and  $a_2\ne a_0^+$, they both must necessarily approach the  point $a_-$ as $t \to +\infty$, as shown in Sec.~\ref{sec:steady},  thus cancelling the perturbation term in (\ref{eq:zbp}). In other words, in the steady regime the solution becomes simply
\begin{align}
z(\zeta, t) =~& 2t +i\frac{\gamma}{2}+ \log \frac{P(e^{-i\gamma}\zeta;\rho^2)}{P(\zeta;\rho^2)},
 \label{eq:zU2}
\end{align}
which corresponds precisely to a steady solution with velocity $U=2$. 
This result, together with the discussion in Sec.~\ref{sec:steady}, demonstrates that the only {\it attractor} of the dynamical system corresponding to the motion of singularities $\{a_k(t)\}$, for $k=1,...,N$,  corresponds to the solution with $U=2$.

It is perhaps instructive to analyze how the selection scenario described above takes place in the physical $z$-plane.  As already noted, the  perturbation enacted by the last term in (\ref{eq:zbp}) splits the pole of ${\cal S}(z,t)$  at infinity into two logarithmic singularities located at some finite (albeit faraway) points. Now, given that the singularities of ${\cal S}(z,t)$ remain fixed in time, they will all eventually be left infinitely far behind the bubble for $t\to\infty$, since we assume that the solution exists for all times. The final result then is that the bubble must reach an asymptotic shape  whose Schwarz function is regular in the entire fluid domain---the only such possibility being a solution with $U=2$.

\begin{figure}[t]
	\center \subfigure[\label{fig:pertub1a}]{\includegraphics[width=0.4\textwidth]{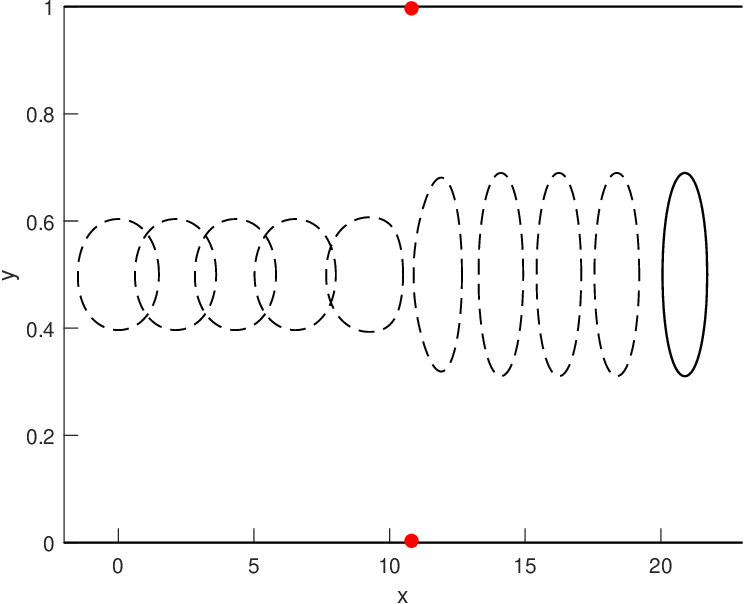}}\qquad
		 \subfigure[\label{fig:pertub1b}]{\includegraphics[width=0.4\textwidth]{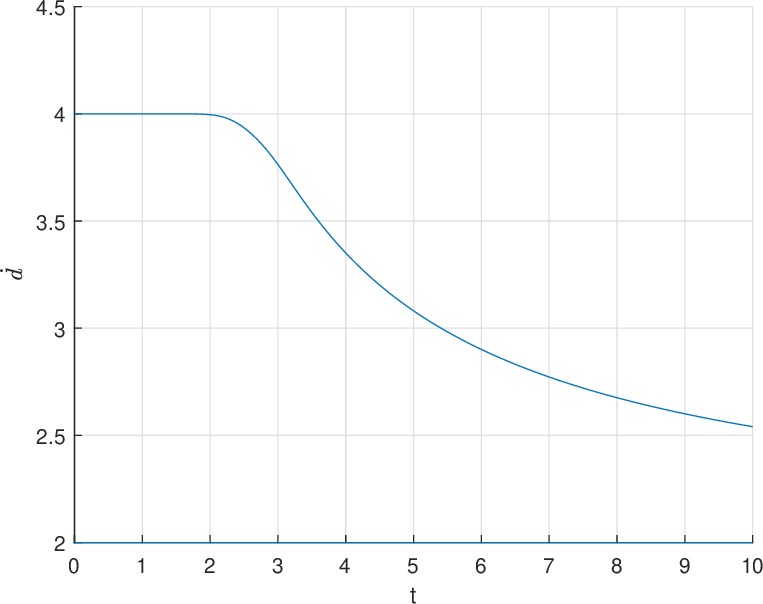}}
		 \subfigure[\label{fig:pertub1c}]{\includegraphics[width=0.4\textwidth]{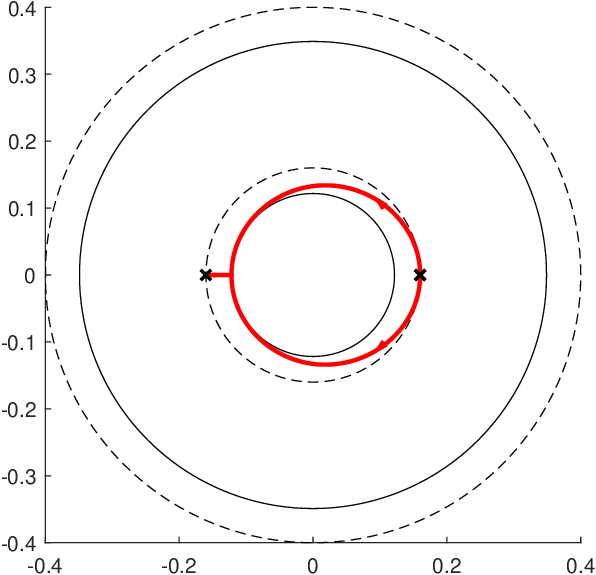}}
				\caption{Shape evolution (b) and velocity (b) after perturbing a steady solution with $U=4$. Here the initial shape corresponds to Eq.~(\ref{eq:zbp}) with parameters $\alpha_0=0$, $\alpha_1=0.25$, $\alpha_2=-0.125$, $\alpha_3=-0.125$, $a_1(0) = -0.16-10^{-4}$, $a_2(0) = 0.16+10^{-7}+10^{-5}i$, $a_3(0) = 0.16+10^{-7}-10^{-5}i$. Time instant shown are $t=0, 0.5274, 1.0793, 1.6306, 2.2904, 3.2246, 4.3490, 5.4190, 6.4867, 7.7407$. (c) Trajectories of the singularities of the map $z(\zeta,t)$  in the $\zeta$ plane.}
	\label{fig:pertub1}
\end{figure}

In Fig.~\ref{fig:pertub1} we show an example of a bubble evolution resulting from a small  perturbation of a symmetric steady shape with $U=4$. In this figure, the perturbation was chosen so that the pole of the Schwarz function at infinity was replaced with three finite log-singularities: two of which were placed on the channel walls (red dots), while the third one lies far behind the bubble (not shown in the figure). At first,  the bubble hardly feels the presence of these new singularities and so it moves, almost without deformation, with a nearly constant  velocity  equal to that of the unperturbed shape (i.e., $U=4$); see Fig.~\ref{fig:pertub1b}. However, as soon as the bubble passes the singularities of ${\cal S}(z,t)$ upstream, it suddenly expands in the transverse direction, thus reducing its speed towards the selected velocity $U=2$. In Fig.~\ref{fig:pertub1c} we show the trajectories (red curves) of the singularities of $z(\zeta,t)$, where one clearly sees that all singularities move towards the point $a_0^-=-\rho^2$. Here, as before, the initial locations of the singularities are indicted by black crosses. Note that the singularity initially at $a_0^+=\rho^2$ has been split into  two nearby singularities that appear as one in the scale of Fig.~\ref{fig:pertub1c}; but for later times they both move apart from $a_0^+$  and toward $a_0^-$, which is the sole attractor of the dynamics.

The preceding stability argument thus shows that, even if we could prepare the real system with an initial condition corresponding to a steady solution with $U\ne 2$, inevitable perturbations will lead to a steady regime where the bubble moves with speed $U=2$. In this context, surface tension is just one of many possible perturbations that can force the system toward the attractor, while also regularizing high curvatures.

\section{Conclusions}
\label{sec:conclusions}

We have presented a general class of analytical solutions for the unsteady motion of a bubble in a Hele-Shaw channel. The solutions are given in terms of a conformal mapping from an annulus to the fluid region outside the bubble, with the corresponding mapping function being written explicitly in terms of certain special (elliptic) functions. The time-dependent parameters that enter the solutions are given implicitly by a set of conservation laws which reflect the integrable nature of the Laplacian growth model. Numerically solving the set of ordinary differential equations that derives from these conservation laws allows us to follow the motion of the bubble in time for any given initial shape.  Several examples of  time-evolving bubbles have been presented. We have  shown, both analytically and numerically, that in the asymptotic limit of large times, i.e., for $t\to\infty$, our time-dependent solutions approach the known steady solutions for a single bubble in a Hele-Shaw  channel \cite{TS,tanveer87, us2014}. 

We have also shown that the steady solutions with $U=2V$ are the only attractor of the dynamics. The analysis reported here confirms in greater detail the argument, put forth in \cite{98PRL} for a finger and in \cite{us2014} for a bubble, that velocity selection in Hele-Shaw interface flows can be explained entirely within the context of the zero surface tension dynamics (Laplacian growth).
In this scenario, the selection mechanism is encoded in the very dynamics of the idealized Laplacian growth model (without any regularizing boundary condition), as the selected pattern is the only attractor of the dynamics.

\begin{acknowledgments}
This work was supported in part by the Conselho Nacional de Desenvolvimento Cient\'ifico e Tecnol\'ogico (CNPq) under Grant Number 307385/2023-0 (GLV).
 \end{acknowledgments}

\appendix


\clearpage

\begin{thebibliography}{99}

\bibitem{Lamb} H. Lamb, {\it Hydrodynamics} (Cambridge University Press, Cambridge, 1906).

\bibitem{Darcy} H. Darcy, {\it Les fontaines publiques de la ville de Dijon} (Dalmont, Paris, 1856).

\bibitem{Porous} S. Whitaker, 
Transp.  Porous Med. {\bf 1}, 3 (1986). 

\bibitem{Langer} J. S. Langer, Rev. Mod. Phys. \textbf{52}, 1 (1980).

\bibitem{combustion} A. P Aldushin, B. J. Matkowsky, Combustion Sci. and Tech. {\bf 133}, 293 (1998).

\bibitem{void} M. Ben Amar, Physica D (Amsterdam, Neth.) {\bf 134}, 275 (1999).

\bibitem{streamer} A. Luque, F. Brau, and U. Ebert,  Phys. Rev. E {\bf 78}, 016206 (2008).

\bibitem{bacteria} J. M\"uller and W. van Saarloos, Phys. Rev. E {\bf 65}, 061111 (2002).

\bibitem{Pelce} P. Pelc\'e, {\it Dynamics of Curved Fronts} (Academic Press,
San Diego, 1988).

\bibitem{MWZ} M. Mineev-Weinstein, P. B. Wiegmann, and A. Zabrodin, Phys. Rev.
Lett. \textbf{84}, 5106 (2000).

\bibitem{KMWZ} I. Krichever, M. Mineev-Weinstein, P. Wiegmann, and A. Zabrodin, 
Physica D {\bf 198}, 1 (2004). 

\bibitem{MPT} M. Mineev-Weinstein, M. Putinar, R. Teodorescu, J. Phys. A {\bf 41}, 263001 (2008).

\bibitem{ABWZ} O. Agam, E. Bettelheim, P. Wiegmann, and A. Zabrodin, Phys. Rev. Lett. {\bf 88}, 236801 (2002).

\bibitem{vasiliev} B. Gustafsson, R. Teodorescu, and A. Vasil'ev,  {\it  Classical and Stochastic Laplacian Growth}  (Birkh\"auser, Basel, 2014).

\bibitem{ST} P. G. Saffman and G. I. Taylor, Proc. R. Soc. Lond. A \textbf{245}, 312  (1958).

\bibitem{SurfaceTension} B. I. Shraiman, Phys. Rev. Lett. \textbf{56}, 2028
(1986); D. C. Hong and J. S. Langer, {\it ibid} \textbf{56},
2032 (1986); R. Combescot, T. Dombre, V. Hakim, Y. Pomeau, and A. Pumir, {\it ibid} \textbf{56}, 2036
(1986); S. Tanveer, Phys. Fluids \textbf{30}, 1589 (1987).

\bibitem{tanveer86} S. Tanveer, Phys. Fluids \textbf{29}, 3537 (1986).

\bibitem{tanveer87} S. Tanveer, Phys. Fluids \textbf{30}, 651 (1987).

\bibitem{TS}  G. I. Taylor and P. G. Saffman, Q. J. Mech. Appl. Maths \textbf{12}, 265 (1959).

\bibitem{GLV2015} G. L. Vasconcelos,  J. Fluid Mech.  {\bf 780}, 299 (2015).

\bibitem{Saffman1959} P. G. Saffman,  
 {Q. J. Mech. Appl. Maths} \textbf{12}, 146--150 (1959).

\bibitem {Howison} S. D. Howison,   {J. Fluid Mech.} {\bf 167}, 439--453 (1986).
 
\bibitem{Mineev94} M. Mineev-Weinstein and S. P. Dawson, Phys. Rev. E {\bf 50}, R24 (1994); Physica D {\bf 73}, 373 (1994).

\bibitem{BensimonPelce}  {D. Bensimon and P. Pelc\'e},  {Phys. Rev. A} \textbf{33}, 4477--4478 (1986).

\bibitem{ShraimanBensimon} {B. Shraiman and D. Bensimon},
{Phys. Rev. A} \textbf{30}, 2840--2842 (1984).

\bibitem{MarkSilvina}
   M. Mineev-Weinstein and S. P. Dawson, {Phys. Rev. E} \textbf{57}, 3063--3072 (1998).
   
\bibitem{98PRL} M. Mineev-Weinstein, Phys. Rev. Lett. \textbf{80}, 2113 (1998).

\bibitem{us2014}  G. L. Vasconcelos and M. Mineev-Weinstein,  Phys. Rev. E  {\bf 89}, 061003(R) (2014).


\bibitem{Robb2015} A. H. Khalid, N. R. McDonald,   and J. M. Vanden-Broeck,    Phys. Fluids {\bf 27}, 012102 (2015).

\bibitem{Green2017}C. C. Green, C. J. Lustri, and S. W. McCue,  Proc. R. Soc. A {\bf 473}  20170050 (2017).

\bibitem{PhysicaD2023} M. Mineev-Weinstein and G. L. Vasconcelos, 	
 Physica D {\bf 459}, 134032 (2024).


\bibitem{king} S. J. Chapman and J. R. King, J. Eng. Math. {\bf 46}, 1-32 (2003). 

\bibitem{mccue}  M. Dallaston and S. McCue, ANZIAM J. {\bf 52}, C124-C138 (2011).

\bibitem{RMP1986} D. Bensimon, L. P. Kadanoff, S. Liang, B. I. Shraiman, and C. Tang, Rev. Mod. Phys. {\bf 58}, 977 (1986).

\bibitem{hadamard} J. Hadamard, Sur les problèmes aux dérivées partielles
et leur signification physique, Princeton University Bulletin {\bf 13}, 49-52 (1902).

\bibitem{lavrentev} M. M. Lavrent'ev, V. G. Romanov, and S. P. Shishatskii, {\it Ill-Posed Problems of Mathematical Physics and Analysis}, Series: Translations of Mathematical Monographs, Vol. 64 (American Mathematical Society, 1986).

\bibitem{tikhanov} A. N. Tikhonov, On the stability of inverse problems, Doklady Acad. Sci. USSR {\bf 39}, 176--179 (1943); On the solution of ill-posed problems and the method of regularization, Dokl. Akad. Nauk SSSR {\bf 151}, 501--504 (1963) (in Russian); On the regularization of ill-posed problems. Dokl. Akad. Nauk SSSR {\bf 153},  49--52 (1963) (in Russian).


\bibitem{Davis} P. J. Davis, \emph{The Schwarz Function and its Applications}, Carus Mathematical Monograph No. 17 (The Mathematical Association of America, 1974).  

\bibitem{Richardson1994} {S. Richardson},
{Eur. J. Appl. Math.} {\bf 5}, 97--122 (1994).

\bibitem{Richardson1996a} {S. Richardson, S.}   
{Phil. Trans. R. Soc. Lond. A} {\bf 354}, 2513--2553 (1996).

\bibitem{pof2012}  {M. C. Dallaston and S. W. McCue}, {Phys. Fluids}  {\bf 24}, 052101 (2012).

\bibitem {Crowdy2002} D. G. Crowdy, 
{Q. Appl. Math.} {\bf 60}, 11--36 (2002).

\bibitem{Jonathan2016c} 	J. S. Marshall,
IMA J. App. Math.  {\bf 81}, 723--749  (2016).

\bibitem{Richardson2001}  S. Richardson,   
{ Eur. J. Appl. Math.} {\bf 12}, 571--599 (2001).

\bibitem{Richardson1996b} S. Richardson, 
Eur. J. Appl. Math. {\bf 7}, 345--366 (1996).

\bibitem{CrowdyTanveer} D. Crowdy and S. Tanveer, 
{J. Stat. Phys.} {\bf 114}, 1501--1536 (2004). 

\bibitem{Jonathan2016a} J. S. Marshall, 
{Q. Jl. Mech. Appl. Math.}  {\bf 69}, 35-66 (2016).

\bibitem{Jonathan2016b} J. S. Marshall,
{Q. Jl. Mech. Appl. Math.}  {\bf 69}, 1--33 (2016).





\bibitem{Shraiman1984} B. Shraiman and D. Bensimon, Phys. Scr. {\bf T9},  123--125 (1985).

\bibitem{Saarlos1988} W. Van Saarlos,  Phys. Rev. A  {\bf 37}, 211--229 (1988).

\bibitem{Saarlos2003} W. Van Saarlos,  Phys. Rep.  {\bf 386}, 29--222 (2003).



\bibitem{Tanveer1990}  S. Tanveer, 
Proc. R. Soc. Lond. A {\bf 428}, 511–545 (1990).


\bibitem{Howison92} S. D. Howison,  Eur. J. Appl. Math. {\bf 3}, 209 (1992). 

\bibitem{GLV2001} G. L. Vasconcelos, J. Fluid Mech. \textbf{444},  175 (2001).

\bibitem{Gradshtein} I. S. Gradshteyn and I. M. Ryzhik, {\it Table of Integrals, Functions, and Products} (Acad. Press, London, 1980).






\end{thebibliography}
\end{document}